\documentclass[letterpaper]{jpconf}
\usepackage{graphicx} 
\usepackage{color}
\usepackage{epsfig}
\usepackage{amsmath,amsfonts,bm,color}

\usepackage[active]{srcltx}
\usepackage{hyperref}

\usepackage{amssymb}

\newcommand{\nlsim}{\mathrel{\rlap{\lower4pt\hbox{\hskip0pt$\sim$}} 
 \raise1pt\hbox{$<$}}}           
\newcommand{\ngsim}{\mathrel{\rlap{\lower4pt\hbox{\hskip0pt$\sim$}} 
 \raise1pt\hbox{$>$}}}

\def\gtorder{\mathrel{\raise.3ex\hbox{$>$}\mkern-14mu
 \lower0.6ex\hbox{$\sim$}}}
\def\ltorder{\mathrel{\raise.3ex\hbox{$<$}\mkern-14mu
 \lower0.6ex\hbox{$\sim$}}}

\def\q2{$Q^2$}

\def\gev2{GeV$^2$}
\def\3he{$^3$He}
\def\GeV2{\,{\rm GeV}^2}
\def\phigg{\phi_{\gamma\gamma}}

\def\xBj{x_{\rm Bj}}

\def\eg{{\em e.g.}}
\def\etal{{\em et al.}}
\def\Real{\Re{\rm e}}
\def\Imag{\Im{\rm m}}
\def\phiGG{\phi_{\gamma\gamma}}

\begin{document}

\begin{flushright}  JLAB-THY-11-8   \end{flushright}

\title[Generalized Parton Distributions]{Deeply Virtual Exclusive Processes and Generalized Parton
Distributions
}

\author{Charles E. Hyde$^{1,2}$, Michel Guidal$^3$, and Anatoly V. Radyushkin$^{2,4,5}$}
\address{$^1$ Laboratoire de Physique Corpusculaire, 
Universit\'e Blaise Pascal, 63177 Aubi\`ere, FRANCE}
\address{$^2$ Department of Physics, Old Dominion University, Norfolk VA, 23529, USA}
\address{$^3$ Institut de Physique Nucl\'eaire ORSAY, 91406 Orsay, France}
\address{$^4$ Thomas Jefferson National Accelerator Facility, Newport News, VA,23606, USA}
\address{$^5$ Bogoliubov Laboratory  of Theoretical Physics, 141980 Dubna, Russia}

\begin{abstract}

The goal of the  comprehensive program in Deeply Virtual Exclusive Scattering at Jefferson Laboratory 
 is to create  transverse spatial images of quarks and gluons as a function
of their longitudinal momentum fraction in the proton, the neutron, and in nuclei.  These functions are the
Generalized Parton Distributions (GPDs) of the target nucleus.  Cross section measurements of the
Deeply Virtual Compton Scattering (DVCS) reaction $ep\rightarrow ep\gamma$ in Hall A support
the QCD factorization of the scattering amplitude 
 for $Q^2\ge 2$ GeV$^2$.  Quasi-free neutron-DVCS measurements on the Deuteron indicate sensitivity to the quark angular momentum
 sum rule.   Fully exclusive H$(e,e' p\gamma )$ measurements
have been made in a wide kinematic range in CLAS with polarized beam, and with
both unpolarized and longitudinally polarized targets. Existing models are qualitatively consistent with  the
JLab data, but there is a clear need for less constrained models.
Deeply virtual vector meson production is studied in CLAS.  The 12 GeV upgrade will be essential for
for these channels. The 
$\rho$ and $\omega$ channels  reactions offer the prospect of flavor sensitivity to the quark GPDs, while 
the $\phi$-production channel is dominated by  the gluon distribution.  

\end{abstract}

\section{Deeply Virtual Exclusive Scattering}

\begin{figure}[b]
\hfill\includegraphics[width=0.8\textwidth]{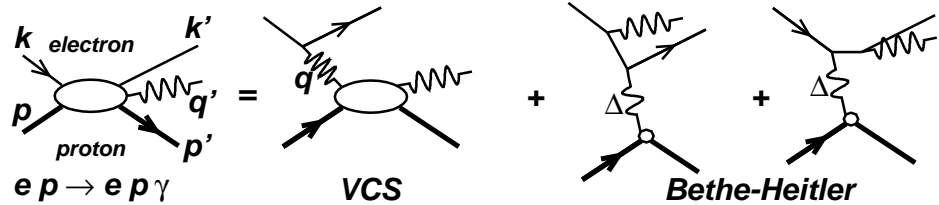}
\caption{\label{fig:FVCS}
Lowest order QED amplitude for the $ep\rightarrow ep\gamma$ reaction.
 The momentum four-vectors of all external particles are labeled at left.
The net four-momentum transfer to the proton is 
$\Delta_\mu=(q-q')_\mu=(p'-p)_\mu$. In the virtual Compton scattering
(VCS) amplitude, the (spacelike) virtuality of the incident photon is
$Q^2=-q^2=-(k-k')^2$.  In the Bethe-Heitler (BH) amplitude, the virtuality
of the incident photon is $-\Delta^2=-t$.  Standard $(e,e')$ invariants
are $s_e=(k+p)^2$, $\xBj=Q^2/(2q\cdot p)$ and $W^2=(q+p)^2$.
}
\end{figure}

In the past decade, Deep Exclusive Scattering (DES) has emerged as
a powerful new probe of the partonic structure of the nucleon,   hadrons
and nuclei.  These are reactions of the type:
\begin{eqnarray}
e+ {
   \rm Nucleon} &\rightarrow& e +{
   \rm Nucleon}  + \gamma 
\nonumber \\
&\rightarrow& e + {
 \rm Nucleon} + {\rm meson}
\label{eq:DVES}
\end{eqnarray}
The 
{\color{black}   $eN\rightarrow eN\gamma$}
reaction 
is the coherent sum of the Bethe-Heitler and virtual Compton amplitudes,
as illustrated in  Fig.~\ref{fig:FVCS}.
The electron scattering kinematics of deeply virtual processes
corresponds to the Deep Inelastic Scattering
(DIS), or Bjorken limit of inclusive electron scattering,
with   $Q^2$ large  and $W^2$  above the resonance region.  In addition,
the forward exclusive limit is defined kinematically by 
\mbox{\color{black}   $-t \lesssim \Lambda_{\rm  QCD}^2 \lesssim 1$ GeV$^2$.}
Thus the deeply virtual Compton scattering (DVCS) amplitude, 
{\color{black}   $\gamma^* N \rightarrow \gamma N$}
is an 
``off-forward'' generalization of the forward Compton amplitude which
defines the DIS cross section via the optical theorem.

\begin{figure}
\hfill\includegraphics[width=0.8\textwidth]{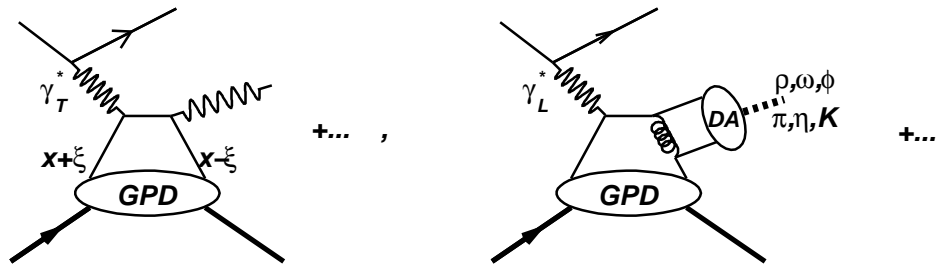}
\caption{\label{fig:DVCS-GPD} 
Factorization of the $\gamma^* p \rightarrow \gamma p$ DVCS amplitude 
and
the $\gamma^* p \rightarrow M N$ deep virtual meson production amplitude 
in the Bjorken limit
of large $Q^2$ and $-t\ll Q^2$.  All permutations of the photon and gluon vertices 
{\color{black}   should be}
included in the amplitudes.  The labels on the quark lines are the light-cone
momentum fractions, relative to $P^+=(p+p')^+/2$.  
}
\end{figure}

{\color{black}   The intense interest in DVCS  started after  the   article 
by Ji \cite{Ji:1996ek}, linking DVCS to the
total contribution of quarks to the proton spin.}
{\color{black}   It was found 
\cite{Ji:1996ek,Radyushkin:1996nd,Radyushkin:1996ru,Ji:1996nm}{} that, 
in analogy with DIS,  in the limit of large $Q^2$ and small $t$, 
amplitudes of DVCS and  deeply virtual meson production (DVMP)
can be expressed in a power series of $1/Q^2$,
with the power determined by twist  (dimension
minus spin)  of each operator in the   expansion. 
Detailed proofs of factorization for DVCS and DVMP were given in 
\cite{Collins:1996fb, Radyushkin:1997ki, Ji:1998xh,Collins:1998be}.}
This factorization is depicted in Fig.~\ref{fig:DVCS-GPD} for the DVCS
and DVMP amplitudes.
In these processes, the leading 
{\color{black}   power term of the}
amplitude is the convolution
of the perturbative kernel with a new class of non-local bi-linear
twist-2 quark (or gluon) operators, called Generalized Parton Distributions (GPDs)
\cite{Ji:1996ek,Radyushkin:1996nd}.
These were first described by M\"uller {\it et al.} \cite{Mueller:1998fv}.
In the case of deeply virtual meson production, the hard kernel is also
convoluted with the meson Distribution Amplitude (DA).

\subsection{Generalized Parton Distributions}

The Generalized Parton Distributions (GPDs) 
{\color{black}   parameterize}  Fourier transforms of 
nucleon matrix elements
of bilinear quark (and gluon)
operators separated by a light-like interval \mbox{$z^2=0$ \cite{Mueller:1998fv}.}
The kinematics are commonly
defined in terms of symmetric variables
(see Figs.  \ref{fig:FVCS} and \ref{fig:DVCS-GPD}) :
\begin{align}
P_\mu &=(p_\mu+p_\mu^\prime)/2, \qquad
\overline{q} = (q_\mu+q_\mu^\prime )/2
\nonumber \\
\xi &= \frac{-\overline{q}^2}{2\overline{q}\cdot P}
         \longrightarrow
         \frac{\xBj}{2-\xBj}\qquad{\rm as}\quad 
{\color{black}   t/Q^2\rightarrow 0} .
\label{eq:xiKin}
\end{align}
The generalized Bjorken variable $\xi$ has the same form with respect to the
symmetrized variables $P$ and $\overline{q}$ as does $\xBj$ with respect to the DIS variables
$p$ and $q$.

It is convenient to 
{\color{black}   use}
a reference frame in which $P^\mu$ has only time- and $z$-components,
both positive.  We define light-cone vectors
\begin{equation}
n^\mu = \left[1,0,0,-1\right]/(\sqrt{2}P^+)\, ,\qquad
\tilde{p}^\mu = \left[1,0,0,1\right]P^+/(\sqrt{2})
\label{eq:lightcone}
\end{equation}
Then in the forward limit of either DVCS or deeply virtual production of
a light meson, 
$-2\xi$ is the   ``$+$'' fraction of both the momentum transfer to the target and the virtual photon:
\begin{eqnarray}
\Delta^+ &= \Delta\cdot n  \approx  -2\xi P^+ \approx q\cdot n = q^+
\label{eq:DeltaPlus}
\end{eqnarray}
The quark GPDs $H$ and $E$ are the nucleon helicity conserving and helicity-flip
 matrix elements of the vector
{\color{black}   operator   containing}
$\gamma\cdot n    = \gamma^+$.
Suppressing the QCD 
scale dependence and the Wilson-line gauge link 
{\color{black}   one   can write the flavor-$f$ dependent GPDs as (see, e.g., } \cite{Diehl:2003ny}):
\begin{align}
\int \frac{dz^-P^+ }{2\pi} e^{ixP^+z^-} &
\left\langle p',s'\right|
\overline{\Psi}_f\left(-z^-/2 
                      \right)\gamma\cdot n \Psi_f\left(z^-/2
                      \right)
\left| p, s \right\rangle \nonumber \\
&=
    \overline{U}(p',s')\left[H_f(x,\xi,t)\,  \gamma\cdot n
    + E_f(x,\xi,t)  \, 
    \frac{i}{2M}n_\alpha\sigma^{\alpha\beta}\Delta_\beta
    \right]\, U(p,s)  \  , 
\label{eq:VectorGPDs}
\end{align}
where the $U(p,s)$ are the nucleon spinors.
The factorization proofs demonstrate that the initial and final $+$ momenta of the
    active parton are $(x\pm \xi)P^+$.
The GPDs $\widetilde{H}$, $\widetilde{E}$ are defined similarly as the matrix elements
of the axial operator containing $n\cdot \gamma\gamma_5 = \gamma^+\gamma_5$:
\begin{align}
\int \frac{ dz^-P^+}{2\pi} e^{ixP^+z^-} &\left\langle p',s'\right|
\overline{\Psi}_f\left(-z^-
                       \right)\gamma\cdot n \gamma_5 \Psi_f\left(z^-
                       \right)
\left| p, s \right\rangle \nonumber \\
  &=
     \overline{U}(p',s') \left[
    \widetilde{H}_f(x,\xi,t)n\cdot \gamma\gamma_5 
     + \widetilde{E}_f(x,\xi,t) \frac{ n\cdot \Delta}{2M}\gamma_5
     \right]U(p,s) \  . 
\label{eq:AxialGPDs}
\end{align}  
With the convention that positive and negative  momentum fractions refer to 
quarks and anti-quarks, respectively, we observe the following kinematic regions of the GPDs
(Fig.~\ref{fig:DVCS-GPD}):
$x>\xi>0$: the initial and final partons are quarks;
$x<-\xi<0$: the initial and final partons are anti-quarks;
$|x|<\xi$: a $q\overline{q}$ pair is exchanged in the $t$-channel.
This identification is reflected in the QCD evolution equations of the GPDs.  
For $|x| > \xi$, the 
{\color{black}   evolution  of GPDs is   similar to  the DGLAP evolution}
of the forward 
parton distributions, whereas for $|x|<\xi$, the GPDs evolve according to the ERBL equations
of a meson DA \cite{Radyushkin:1996nd}.

The GPDs combine the momentum fraction information of the forward parton
distributions (PDFs) of DIS with the transverse spatial information of the elastic
electro-weak
form factors.  The forward limits of the helicity conserving GPDs are
\begin{eqnarray}
\left. {H_f(x,0,0) \atop \widetilde{H}_f(x,0,0) } \right\} =  
\begin{cases} q_f(x)\theta(x)-\overline{q}_f(-x)\theta(-x) 
\\
        \Delta q_f(x)\theta(x)+\Delta\overline{q}_f(-x)\theta(-x)
,
\end{cases}
\label{eq:pdf} 
\end{eqnarray}
where $q_f(x)$ and $\overline{q}_f(x)$ are the   flavor-$f$
 dependent quark and anti-quark
momentum-fraction distributions; 
and  $\Delta q_f(x)$ and $\Delta \overline{q}_f(x)$
are the quark and anti-quark helicity distributions.  
There are no specific analogous constraints on the forward limits
of $E$ and $\widetilde{E}$.

The first moments of the GPDs are equal to the corresponding elastic form factors.
\begin{eqnarray}
  \int dx \left\{{H_f(x,\xi,t) \atop E_f(x,\xi,t)}\right\} = 
\begin{cases}
F_{1,f}(-t) \\
F_{2,f}(-t)
\end{cases}
\label{eq:VectorFF} 
\end{eqnarray}
where 
{\color{black}   $F_{1,f}(-t)$, $F_{2,f}(-t)$ }
are  
{\color{black}   flavor-$f$ components of } 
the Dirac and Pauli form factors 
of the proton, defined with positive arguments
in the space-like regime.  Similarly,
\begin{eqnarray}
\int dx \left\{ {\widetilde{H}_f(x,\xi,t) \atop \widetilde{E}_f(x,\xi,t)}\right\} 
  = 
\begin{cases}
g_{A,f}(-t) 
\\
g_{P,f}(-t)
, 
\end{cases}
\label{eq:AxialFF} 
\end{eqnarray}
where  $g_{A,f}$ and $g_{P,f}$ are the flavor 
{\color{black}   components of the}
axial and pseudoscalar form factors of the proton.   

The $\xi$-independent integrals of (\ref{eq:VectorFF}) and (\ref{eq:AxialFF}) are 
examples of the polynomiality condition required by Lorentz invariance.
Specifically, the $x^N$ moment of a GPD is polynomial in even powers of $\xi$, with
maximal power $\le (N+1)$ for $H$, $E$ and maximal power $\le N$ for 
$\widetilde{H}$, $\widetilde{E}$.  The second moment of the GPD sum $H+E$ leads to the
important angular momentum sum rule of Ji
\cite{Ji:1996ek}:
\begin{eqnarray}
\lim_{t\rightarrow 0}\int_{-1}^{1}dx\, x \left[ H_f(x,\xi,t) + E_f(x,\xi,t)\right] &= 2J_f \, ,
\end{eqnarray}
where $2J_f$ is the fraction of the spin of the proton carried by quarks of flavor $f$,
including both spin and orbital angular momentum.
More generally, at non-zero $t$, the individual second moments are form factors
(i.e., Fourier transforms of spatial distributions) of the nucleon's energy-momentum tensor
\cite{Ji:1996ek}.
\begin{eqnarray}
\int_0^1 dx\, x H_f(x,\xi,t) &=& M_{2,f}(t)+\frac{4}{5}\xi^2d_{1,f}(t)
\nonumber \\
\int_0^1 dx\, x E_f(x,\xi,t) &=& \left[ 2 J_f(t) - M_{2,f}(t) \right] -\frac{4}{5}\xi^2d_{1,f}(t) \  .
\label{eq:SecondMoments}
\end{eqnarray}
The forward limit $M_{2,f}(0)$ is the ordinary momentum sum rule
\begin{align}
M_{2,f}(0) &= \int_0^1 x dx \left[q_f(x)+\overline{q}_f(x) \right] \   .
\label{eq:MomentumSum}
\end{align}
The term $d_{1,f}(t)$ is Fourier conjugate to the spatial distribution of the time-averaged 
shear stress on quarks in the nucleon \cite{Polyakov:2002yz,Belitsky:2005qn}.

The GPDs themselves, and not just the moments, provide unique spatial information about
partons in the hadronic target.  The parton impact parameter {\bf b} is Fourier conjugate
 to $\Delta_\perp$.  The Fourier transform 
of $H_f(x,0,\Delta^2)$
determines a positive-definite probability distribution of quarks of flavor $f$ as a function
of longitudinal momentum fraction $x$ and spatial coordinate ${\bf b}$
 in the transverse plane \cite{Burkardt:2000za,Ralston:2001xs}.  
As a consequence, the Dirac form factors $F_1^{p,n}$ are the 2D Fourier transforms of
the charge densities of the proton and neutron in impact parameter space.  A recent analysis
reveals the presence of a negative charge density at the heart of the neutron\cite{Miller:2007uy}.
In the context of the Double Distribution models of the GPDs (see section \ref{sec:VGG}), 
this can be understood
by the excess of down quarks over up quarks in the neutron at large $x$.
Similarly to $H$, the combination of $H_f$ and $E_f$ at $\xi=0$ determines the
spatial density of quarks in a transversely polarized proton
\cite{Burkardt:2002hr}.  
These distributions strongly
break azimuthal symmetry about the longitudinal axis.  In particular, the centroid
of the up and down distributions are displaced in opposite directions, as required by the
fact that the Lorentz boost of a magnetic dipole produces an electric dipole field.
For $\xi\ne 0$, the Fourier transform of the GPDs
determines  overlap matrix elements for partons of initial and final impact parameter
${\bf b}/(1\pm\xi)$ with respect to the center-of-momentum of the initial and final proton
\cite{Diehl:2002he}.
In the particular $x=\xi$ case, the variable ${\bf r}$, Fourier conjugate to $\Delta_\perp,$ is
the transverse separation of the active parton from the center-of-momentum of
the spectator partons \cite{Burkardt:2007sc}.

The experimental program to establish the domain of factorization in DES and to extract  GPDs
promises new insight into 
the quark-gluon structure of hadrons.
The GPDs offer for the first time a probe of
the  rich correlations between 
spatial and momentum degrees of freedom of quarks and gluons in hadronic systems.

\subsection{Scattering Amplitude and Observables}

The $ep\rightarrow ep\gamma$ cross section has the form
(Fig.~\ref{fig:FVCS})
\begin{eqnarray}
\frac{d^5\sigma}{dQ^2 d\xBj d\phi_e d\Delta^2 d\phi_{\gamma\gamma} }
    =
    \frac{\alpha_{QED}^3}{4(2\pi)^2}\frac{\xBj y^2}{Q^4}
 \frac{1}{\sqrt{1+\epsilon_{DVCS}^2}}
    \left[
    \left| {\mathcal T}_{\rm BH}\right|^2+{\mathcal I}+
 \left|{\mathcal T}_{\rm VCS}\right|^2
    \right]  \   ,
\label{eq:dsigma}
\end{eqnarray}
where $\epsilon_{DVCS}^2 = 4\xBj^2 M^2/Q^2$ and ${\mathcal I}$ is the BH$\cdot$VCS
interference term.  
The pure Bethe-Heitler term $|{\mathcal T}_{\rm BH}|^2$
is exactly calculable in terms of the nucleon form factors 
\cite{Guichon:1998xv,Belitsky:2001ns,Goeke:2001tz}.   The full VCS amplitude is expressed as
\begin{align}
{\mathcal T}_{\rm VCS}(e^\pm) &= \overline{u}(k',\lambda)\gamma_\mu u(k,\lambda)
  \frac{\color{black}   (\pm e)}
{q^2}H^{\mu\nu}  
{\color{black} \epsilon_{\nu}^\dagger}
\  .
\label{eq:TVCS}
\end{align}
The general VCS hadronic tensor $H$ has 12 independent terms. 
\index{}In  the leading order twist-2 approximation, $H$ reduces to just four terms
{\color{black} \cite{Ji:1996ek}}
\begin{eqnarray}
{\color{black}   H_{\rm LO, twist \,  2}^{\mu\nu}} 
=& 
         \frac{1}{2} \left(-g^{\mu\nu}\right)_\perp 
           \overline{U}(p') \left[ (n\cdot \gamma){\mathcal H}(\xi,t)
             +\frac{i}{2M}n_\kappa \sigma^{\kappa\lambda}\Delta_\lambda {\mathcal E}(\xi,t)
           \right] U(p)
\nonumber \\
&- \left(\epsilon^{\mu\nu}\right)_\perp
           \overline{U}(p') \left[ (n\cdot \gamma\gamma_5)\widetilde{\mathcal H}(\xi,t)
                                     +(\gamma_5 n\cdot \Delta ) \widetilde{\mathcal E}(\xi,t)
           \right] U(p) \  , 
\label{eq:Hmunu}
\end{eqnarray}
where the transverse tensors are defined as
\begin{equation}
(-g^{\mu\nu})_\perp =  -g^{\mu\nu}+n^\mu \tilde{p}^\nu+\tilde{p}^\mu n^\nu\, ,
\qquad
(\epsilon^{\mu\nu})_\perp = \epsilon^{\mu\nu\alpha\beta}n_\alpha \tilde{p}_\beta \, .
\label{eq:tensors}
\end{equation}
The Compton form factors (CFF) ${\mathcal H}\ldots$ in (\ref{eq:Hmunu})
are defined by the integration over the quark loop in Fig.~\ref{fig:DVCS-GPD}:
\begin{align}
\left[ \mathcal H, \mathcal E\right](\xi,t) &= \int_{-1}^{+1} dx 
               \left[ \frac{1}{x-\xi+i\epsilon}+ \frac{1}{x+\xi-i\epsilon}\right] 
               \left[H, E\right](x,\xi,t)  \   , 
\nonumber \\
\left[\widetilde{\mathcal H}, \widetilde{\mathcal E}\right](\xi,t) &= \int_{-1}^{+1} dx 
               \left[ \frac{1}{x-\xi+i\epsilon}- \frac{1}{x+\xi-i\epsilon}\right] 
               \left[\widetilde{H}, \widetilde{E}\right](x,\xi,t)\, .
\label{eq:HCFF}
\end{align}
Thus the imaginary 
{\color{black}   parts }
of the CFFs are proportional to the GPDs at the point $x=\pm\xi$.
Complete expressions for the VCS hadronic tensor to twist 3 accuracy are 
given in  \eg \cite{Goeke:2001tz,Belitsky:2005qn}.

The importance of the  azimuthal distributions of the DVCS and interference  terms, both
for testing factorization and extracting constraints on the GPDs was first pointed out
by  Diehl \etal  \cite{Diehl:1997bu}.
The azimuthal distribution of the DVCS and interference terms has the
\mbox{ general form
\cite{Belitsky:2001ns}}
\begin{align}
|\mathcal T_{\rm DVCS}|^2 
  &=
  \frac{e^6 (s_e-M^2)^2}{\xBj^2 Q^6}\left\{
  \sum_{n=0}^{2} c_n^{\rm DVCS} \cos(n\phiGG) +  
 \sum_{n=1}^{2}s_n^{\rm DVCS}\sin(n\phiGG)
  \right\} \  ,
\label{eq:sigmaDVCS}\\
\fl
{\mathcal I}(e^\pm)
   &=  \frac{\pm e^6 \xBj^2 (s_e-M^2)^3}{\Delta^2 Q^2 (k-q')^2 (k'+q')^2}
   \left\{
  \sum_{n=0}^{3} c_n^{\mathcal I} \cos(n\phiGG) 
  + \sum_{n=1}^{3}s_n^{\mathcal I}\sin(n\phiGG)
  \right\}  \ .
  \label{eq:sigmaI}
\end{align}
The azimuthal angle $\phiGG$ of the hadronic ${\bf q^\prime}\otimes {\bf p^\prime}$ plane
relative to the electron scattering plane is defined such that $\sin(\phiGG)>0$ when 
$({\bf k}\wedge{\bf k^\prime})\cdot{\bf q^\prime}>0$ and $\phiGG=0$ when the final photon
is on the beam side of ${\bf q}$.

The leading order twist-2 DVCS amplitude couples only to transverse photons.
Consequently, there is a specific twist-hierarchy to the terms in 
(\ref{eq:sigmaDVCS},\ref{eq:sigmaI})
\cite{Belitsky:2001ns}:
\begin{enumerate}
\item 
$c_0^{\rm DVCS}$,  $c_0^{\mathcal I}$ are twist-2;
\label{it:c0}
\item
$(c,s)_1^{\mathcal I}$ are twist-2 (transverse VCS  in interference with longitudinal BH amplitudes);
\item
$(c,s)_1^{\rm DVCS}$ are twist-3 (LT electroproduction interference terms);
\item
$(c,s)_2^{\rm DVCS}$,   are  bilinear combinations of ordinary twist-2 and gluon transversity terms;
\item
$(c,s)_2^{\mathcal I}$ are twist-3;
\item
$(c,s)_3^{\mathcal I}$ are linear in the twist-2 gluon transversity terms.
\label{it:c3}
\end{enumerate}
For example, the $c,s_1^{\mathcal I}$  terms for unpolarized and longitudinally
polarized targets are respectively:
\begin{align}
\left.
\begin{array}{r}
c_{1,\rm unp}^{\mathcal I}  \\
s_{1,\rm unp}^{\mathcal I} 
\end{array}
\right\}
&\propto 
 \left\{
     \begin{array}{c}
     \Real \\ \lambda \Imag 
\end{array} \right\} \mathcal C_{\rm unp}^{\mathcal I}
\  , 
\nonumber \\ &
\mathcal C_{\rm unp}^{\mathcal I} =
\left[ 
          F_1 {\mathcal H} + \xi G_M\widetilde{\mathcal H}
          + \tau_C F_2 {\mathcal E}
          \right]  \   , 
\label{eq:CI-unp} \\
\left.
\begin{array}{r}
c_{1,\rm LP}^{\mathcal I}  \\
s_{1,\rm LP}^{\mathcal I} 
\end{array}
\right\}
&\propto  \Lambda
 \left\{
     \begin{array}{c}
    \lambda \Real \\  \Imag 
\end{array} \right\} \mathcal C_{\rm LP}^{\mathcal I} \  , 
\nonumber \\
 & C_{\rm LP}^{\mathcal I} =\left[ 
          \xi G_M \left({\mathcal H}+\frac{\xBj}{2}\mathcal E\right)
          + F_1\widetilde{\mathcal H}
          - \xi \left(\frac{\xBj}{2}F_1+\tau_C F_2\right) \widetilde{\mathcal E}
          \right], 
\label{eq:CI-LP}
\end{align}
where  $\Lambda $ is the target polarization and  $\tau_C = -\Delta^2/(4M^2)$
  \cite{Belitsky:2001ns}.
The EM form factors $F_1$, $F_2$, $G_M$ are evaluated at $-t$ and 
the Compton form factors ${\mathcal H}$, ${\mathcal E}\ldots$ are evaluated at $(\xi,t)$.
In particular, we expect that on the proton the $F_1 {\mathcal H}$ and 
 $F_1 \widetilde{\mathcal H}$ terms will dominate the unpolarized and longitudinal target
 polarization observables of  (\ref{eq:CI-unp}) and (\ref{eq:CI-LP}), respectively.
 On the other hand, on the neutron both $F_1$ and $g_1(x) = \widetilde{H}(x,0,0)$ are small.
 We anticipate a greater sensitivity to $E$ from the interference term on an
 unpolarized neutron  and to the combination of  ${\mathcal H}$ and $\widetilde{\mathcal E}$
 on a longitudinally polarized neutron.
 Transverse target observables depend on different combinations of CFFs \cite{Belitsky:2001ns}.    
Dynamic and kinematic twist-3 terms in the scattering amplitude
cause kinematically suppressed twist-3 terms to mix into the ``twist-2'' observables listed above
\cite{Belitsky:2008bz,Guichon:2009}.  In addition, any 
{\color{black}   ``twist-2''}
observable will naturally contain
contributions of all higher even-twist in a power series in $1/Q^2$.
The precise  twist content of DES observables must be
determined via a \mbox{$Q^2$-dependent} analysis at fixed $(\xi,t)$.  A complete analysis
must also include the logarithmic $Q^2$-dependence from QCD evolution.

\subsection{Models}
\label{sec:VGG}

Several models of the GPDs exist.  To varying degrees, they incorporate
the theoretical and empirical constraints on the GPDs.
In the valence region, the most widely used models are based on the
 Double Distribution (DD) \
{\color{black}   ansatz proposed by Radyushkin \cite{Radyushkin:1998es}. }
Detailed versions of this
 model are presented  by Vanderhaeghen,  Guichon, Guidal (VGG) \cite{Vanderhaeghen:1999xj}
and Goeke, Polyakov, 
Vanderhaeghen \cite{Goeke:2001tz}.  The Double Distributions re-parameterize the $(x,\xi)$
dependence of the GPDs  in terms of the momentum fractions $\beta$ and $\alpha$,
of $P^+$ and $\Delta^+$, respectively.  Thus the initial and final parton $+$ components
of momentum are $\beta P^+ \mp(1\pm \alpha)\Delta^+ /2$.
The $H$, $E$, and $\widetilde{H}$ Double Distributions are parameterized as:
\begin{align}
GPD_{f,DD}(x,\xi,t) &= \int_{-1}^{+1}d\beta  \int_{-1+|\beta|}^{1-|\beta|} d\alpha \,
           \delta(x-\beta-\alpha\xi)F_f(\beta,\alpha,t)
\nonumber \\
F_f(\beta,\alpha,0) &= h(\beta,\alpha)  \left\{
                              \begin{array}{l}
q_f(\beta); \\
\kappa_f q_f(\beta) (1-\beta)^{\eta_f}/A_f ; \\ \Delta q_f(\beta) ;   
\end{array}
\right.
\label{eq:DD}
 \end{align}
for  $H$, $E$, and $\widetilde{H}$, respectively.  In (\ref{eq:DD}) $q_f$ and $\Delta q_f$
are  the ordinary and helicity-dependent parton distribution functions of flavor $f$ 
and $\kappa_f$ is the flavor anomalous magnetic moment of the proton.  
The normalization of $E$ is such that
 \mbox{$A_f = \int d\beta (1-\beta)^{\eta_f}q_f(\beta)$.}
 The profile function $h$ is commonly parameterized as
 \begin{align}
 h(\beta,\alpha) &= \frac{\Gamma(2b+2)}{2^{2b+1} \Gamma(b+1)}
                               \frac{[(1-|\beta|)^2 - \alpha^2)]^b}
{\left( 1 - |\beta|\right)^{2b+1}}
\label{eq:DD_profile}
 \end{align}
 For $b=1$, this form reduces at $\beta=0$ to an 
asymptotic meson DA $\Phi(\alpha)=3(1-\alpha^2)/4$,
 with support $-1<\alpha<1$.  This connection is  suggested by
 the ERBL evolution equations for $|x|<\xi$. In  general, the exponent $b$ is a free parameter
 and for $b\rightarrow\infty$, the GPD is  $\xi$-independent. 

The DD form of (\ref{eq:DD}) ensures that the polynomiality conditions are automatically
satisfied.  However, it was pointed out by Polyakov and Weiss that for $H$ and $E$, an
additional ``$D$-term'' must be included
\cite{Polyakov:1999gs}
{\color{black}   to produce the highest $\xi^{N+1}$ power  for $x^N$ moment.}
  This term, which only has support in the ERBL region,
is an iso-singlet and enters with opposite sign to \mbox{$H$ and $E$:}
\begin{align}
H_f(x,\xi,t) &= H_{f,DD}(x,\xi,t) + \theta\left(\xi-|x|\right)\frac{1}{N_f}D\left(x/\xi,t \right) \  ,
\nonumber \\
E_f(x,\xi,t) &= E_{f,DD}(x,\xi,t) - \theta\left(\xi-|x|\right)\frac{1}{N_f}D\left(x/\xi,t \right) \  .
\end{align}
In practice, the $D$-term has been taken as an expansion in odd Gegenbauer polynomials,
with the first few terms fitted to a Chiral Soliton model calculation
\cite{Petrov:1998kf,Goeke:2001tz}.
 The $t$ dependence is introduced into the model via a Regge inspired ansatz
 \cite{Goeke:2001tz}:
\begin{eqnarray}
F_f(\beta,\alpha,t) &= h(\beta,\alpha) q_f(\beta) |\beta|^{-\alpha' t}  \   .
 \end{eqnarray}
{\color{black}    As} 
found  in \cite{Guidal:2004nd},
  the high-$t$ form factor is dominated by the contribution in the DD at
  large $\beta$, for which  
 the simple Regge forms must be modified
{\color{black}   to  describe the data. }
{\color{black}   A}
fit to all of the nucleon form factors data 
{\color{black}   was  obtained} 
with the
 ansatz
\begin{eqnarray}
F_f(\beta,\alpha,t) &= h(\beta,\alpha) q_f(\beta) |\beta|^{-(1-\beta)\alpha' t}.
 \end{eqnarray}  
 Aside from the choice of $b$-parameter for each flavor and GPD, variants of the model
 exist with different choices of including  the valence or  valence plus sea contributions
 to $E$ and $H$.  
 The GPD $\widetilde{E}$ is generally parameterized separately,
 as the pion-pole in the $t$-channel.  In this framework, $\widetilde{E}$ is directly related
 to the pion form factor\cite{Goeke:2001tz,Penttinen:1999th}.
 
 The family of models sketched above, and generically labeled ``VGG''
  is qualitatively successful in describing
 the DVCS data.  However, the model is highly constrained and does not have the
 full degrees of freedom of the GPDs.  More general parameterizations will be
 needed   as the data   improves in precision  and covers both a broader 
 kinematic range and a more complete set of spin and flavor observables.

Another approach  \cite{Diehl:1998kh,Brodsky:2000xy} is to construct valence 
generalized parton distributions  as an  overlap of  
light-cone wave functions.    However, a   model  involving 
 the lowest Fock state component  only produces GPDs vanishing 
at the border points $x = \pm \xi$ and in the whole central  region
$-\xi <x< \xi$   \cite{Diehl:1998kh,Boffi:2002yy}.   
It was shown \cite{Ji:2006ea} that inclusion of the higher Fock state components 
gives GPDs that are nonzero in the central  region and at the border points.
In particular, one may assume that overlap of the lowest Fock state components
gives  model GPDs at a low normalization point $Q_0 \sim 300$\,MeV, and then evolve
them to hard scales $Q \gtrsim 1$\,GeV: the evolution will induce nonzero values 
for GPDs in the central region.  Originally \cite{Frankfurt:1997ha},   the evolution approach was used 
to build a model for the  gluon GPD, assuming that for a low normalization point
$Q_0$ it coincides with the usual (``forward'') gluon density,
$H^g (x,\xi; Q_0) = x G(x,Q_0)$.  In Ref.~\cite{Freund:2002qf} this ansatz was also applied for quark
distributions  in  an attempt to describe HERA DVCS data at  low $\xBj$, for which predictions 
based on the double distribution ansatz are too large in magnitude. 
More recently, the 
   ``Dual Parameterization'' (DP) framework   developed by 
Polyakov and collaborators \cite{Polyakov:2002wz,Guzey:2005ec}  was 
used to address this issue.   In this approach,  GPDs are expanded in terms of the
 partial waves exchanged in the $t$-channel.  It was expected that for low $\xBj$  of 
 the HERA DVCS data,   
the expansion
may  be truncated to the first ``forward-like'' functions
\cite{Guzey:2005ec}.  However, a detailed analysis \cite{Guzey:2008ys}
demonstrated that the minimal  model of the dual parameterization 
significantly (by a factor of 4) overestimates the HERA data.
The relation between the dual parameterization approach and 
the double distribution ansatz was investigated in   Ref. \cite{Polyakov:2008aa},
where it was shown that GPDs  built from DD-based models with $b=1$ in Eq.~(\ref{eq:DD_profile}) 
and small $\xi$  may be reproduced just by  the first term of the dual parameterization expansion,
i.e.,   the minimal DP- and DD-based models give  similar results   for DVCS at small $\xBj$,
and both give a rather large value $\sim 1.8$ for the ratio 
$R^{\Sigma}(\xi) \equiv H^{\Sigma}(\xi,\xi)/\Sigma(\xi)$ of singlet quark distributions 
 for small $\xi$, while experimental data  favor 
the value close to 1. In a model developed by D. Mueller and collaborators \cite{Kumericki:2009uq}
it is possible to keep the value of $R^{\Sigma}(\xi)$ ``flexible'', i.e.  to adjust it  to  
describe the data. The ``flexibility'' may be achieved also in the dual parametrization 
approach, if one adds the second ``forward-like''  function.

\section{Initial DVCS Experiments}
\label{sec:initDVCS}

The richness of physical information in the GPDs has sparked an intense experimental
effort.  The H1, HERMES, and CLAS collaborations published the first evidence 
for the DVCS reaction  in 2001.

\subsection{DVCS at HERA}
\label{subsec:DVCS-HERA}
The
H1~\cite{Adloff:2001cn,Aktas:2005ty,Aaron:2007cz} and 
ZEUS~\cite{Chekanov:2003ya,Chekanov:2008vy}
collaborations measured the $p(e,e'\gamma)X$ cross section,
 integrated over $\phi_{\gamma\gamma}$.  The exclusive $p(e,e'\gamma)p$
 channel is enhanced over $p(e,e'\gamma)N^\ast$ channels by  
 vetoing on  forward detectors \cite{Aaron:2007cz,Chekanov:2008vy}.
 In ZEUS, a subset of  $p(e,e'\gamma p)$ events were tagged in a forward tracker
 \cite{Chekanov:2008vy}.
 The HERA data cover a wide kinematic range at low $\xBj$, with central values
 of $Q^2$ and $W$ from 8 to 85 GeV$^2$ and  45 to 130 GeV, respectively.

The HERMES collaboration measured  the $\phi_{\gamma\gamma}$-distribution of the
relative 
beam-helicity asymmetry in the H$(\vec{e},e'\gamma)X$ reaction at average kinematics
$\langle Q^2, \,\xBj,\, t\rangle = (2.6\,{\rm GeV}^2,\, 0.11,\, -0.27\,{\rm GeV}^2)$ \cite{Airapetian:2001yk}.
The FWHM of the $M_X^2$ distribution was $\approx 1\,{\rm GeV}^2$, therefore covering the
majority of the resonance region.  However, at low $\xBj$ and $-t$, model estimates
indicate that the exclusive H$(\vec{e},e'\gamma)p$ channel is dominant\cite{Airapetian:2001yk}.
The HERMES collaboration has recently measured  the beam-charge
asymmetry \cite{Airapetian:2006zr},  transversely polarized target asymmetries
\cite{:2008jga}, longitudinally polarized target asymmetries \cite{:2010mb}, and a more extensive
set of beam spin asymmetries \cite{:2009rj}.
 The final 2006-2007 HERMES run utilized a new recoil detector \cite{Seitz:2004kw}, to establish
exclusivity via H$(\vec{e},e'\gamma p)$ triple coincidence \cite{Lehmann:1900zz}.

\subsection{Initial CLAS DVCS Data}

\label{subsec:initCLAS}

\begin{figure}
  \includegraphics[width=0.49\textwidth]{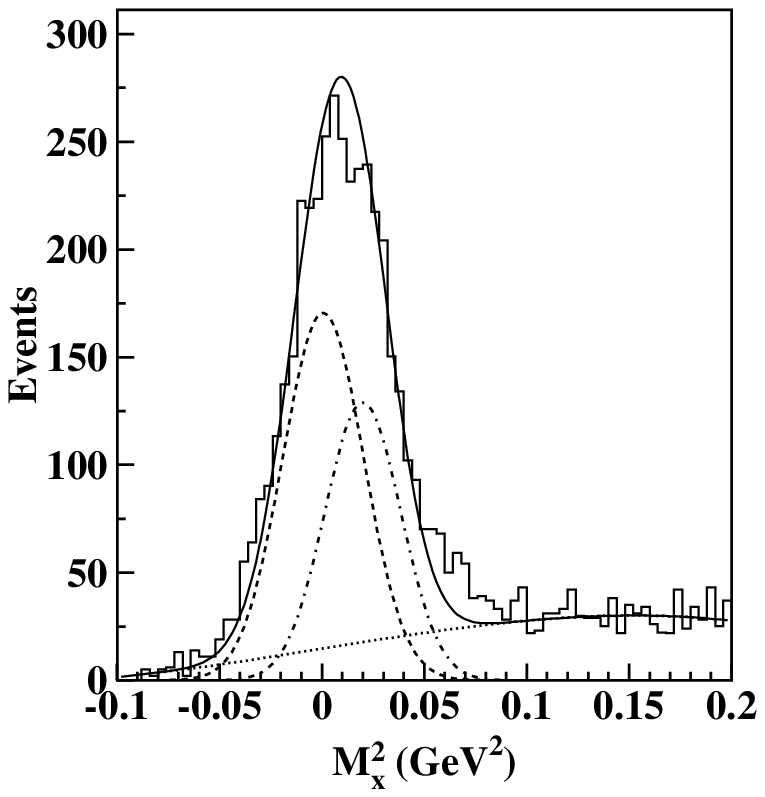}
 \includegraphics[width=0.49\textwidth]{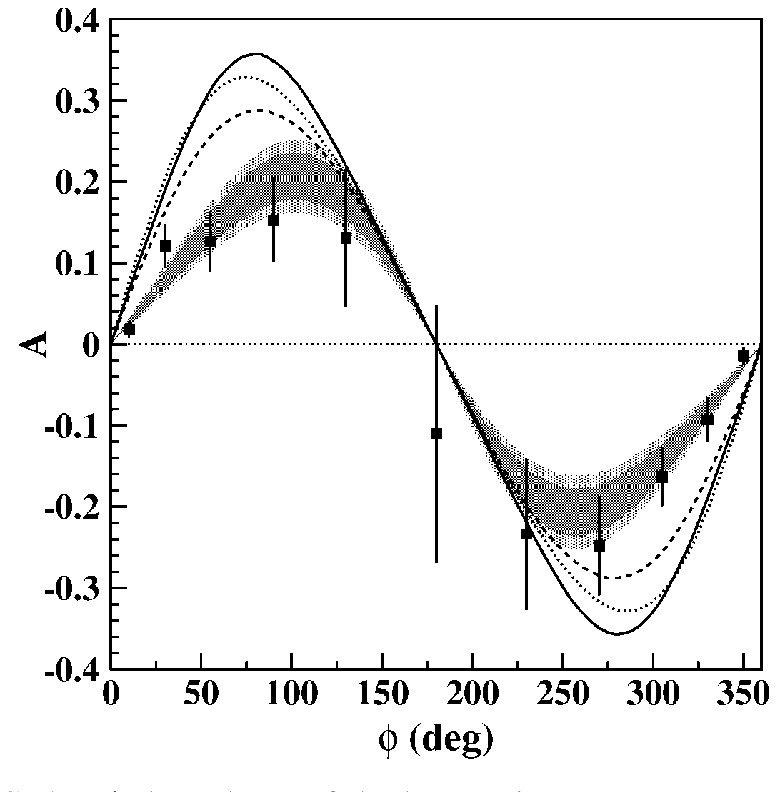} \\
\caption{\label{fig:CLAS2001-BSA}
H$(\vec{e},e'p)x$ analysis from  CLAS at 4.25 GeV incident
electron energy \cite{Stepanyan:2001sm}. 
 \mbox{{\bf Left:} Missing} mass squared $M_x^2$ distribution.  
 The two Gaussian distributions representing the
 H$(\vec{e},e'p)\gamma$ and  H$(\vec{e},e'p)\pi^0$ 
events are described in the text.
 The smooth polynomial background represents the contribution of processes such
 as $ep\rightarrow eN^*\gamma$.
 {\bf Right:}  DVCS Beam-helicity asymmetry.   The  kinematics are integrated over 
$ Q^2\in [1.,\,1.75]\,{\rm GeV}^2$ and $-t\in [0.1,\,0.3]\,{\rm GeV}^2$.
The shaded region is the fit described in the text,
 in section \ref{subsec:initCLAS}.
The curves, described in the text, are evaluated at the fixed values
$Q^2 = 1.25\,{\rm GeV}^2$, $\xBj = 0.19$, and $t=-0.19\,{\rm GeV}^2$. 
}
 \end{figure}

The JLab CLAS Collaboration first 
measured the relative
beam-helicity asymmetry in the H$(\vec{e},e'p)x$ reaction 
with 4.25 GeV incident electrons \cite{Stepanyan:2001sm}.  The exclusive photon was detected
in only a fraction of the acceptance, due to the limited small angle acceptance of the standard CLAS calorimeter.  The $M_x^2$ distribution is shown in Fig.~\ref{fig:CLAS2001-BSA} (left).
  The position and width of the exclusive 
 H$(\vec{e},e'p)\gamma$ event distribution was constrained to fit a subsample of 
  H$(\vec{e},e'p\gamma)$ data at small $(e'\gamma)$ opening angle, such that the events
  are dominated by the BH process.  Similarly, exclusive   $\pi^0$ event distributions  were 
  constrained to a subset of H$(\vec{e},e'p\gamma\gamma )$ events from $\pi^0$ decay.
  Thus the H$(\vec{e},e'p)x$ events in the exclusive region were fitted with two gaussians, for
  the $x=\gamma$ and $x=\pi^0$ channels.  The widths and positions
  of these two gaussians are {\em a priori} constrained.  In this way the exclusive H$(\vec{e},e'p)\gamma$
  channel was isolated.
  The resulting DVCS beam-helicity asymmetry is shown in Fig.~\ref{fig:CLAS2001-BSA} (right).
  The shaded band is a one-sigma fit of the form
  \begin{eqnarray}
  {\rm BSA} &= \alpha \sin(\phi_{\gamma\gamma}) + \beta \sin(2\phi_{\gamma\gamma})
 \label{eq:BSA-CLAS1} 
  \end{eqnarray}
  The $\alpha$ coefficient contains the twist-2 
  physics.
  The $\beta$ coefficient contains the twist-3 physics, as well as contributions from
  $\cos(\phi_{\gamma\gamma}) $ terms in the unpolarized cross section in the denominator
  of the beam spin asymmetry.
  The dashed and dotted curves in  Fig.~\ref{fig:CLAS2001-BSA} are  leading twist calculation of
  the VGG model, in 
  the $\xi$-independent (at fixed $x$) and  $\xi$-dependent
  versions, respectively \cite{Vanderhaeghen:1999xj}.  The solid curve includes an estimate
  of twist-3 effects \cite{Kivel:2000fg,Belitsky:2001yp}.  The models, though constrained 
  by fundamental principles are still  very preliminary.
  It is remarkable that the data and models are in as good agreement
  as indicated by Fig. \ref{fig:CLAS2001-BSA}.  Within the VGG model, the largest contribution
  to the beam helicity asymmetry on the proton comes from the $H(\pm\xi,\xi,t)$ GPD.

\begin{figure}
\centerline{\includegraphics[width=0.75\textwidth]{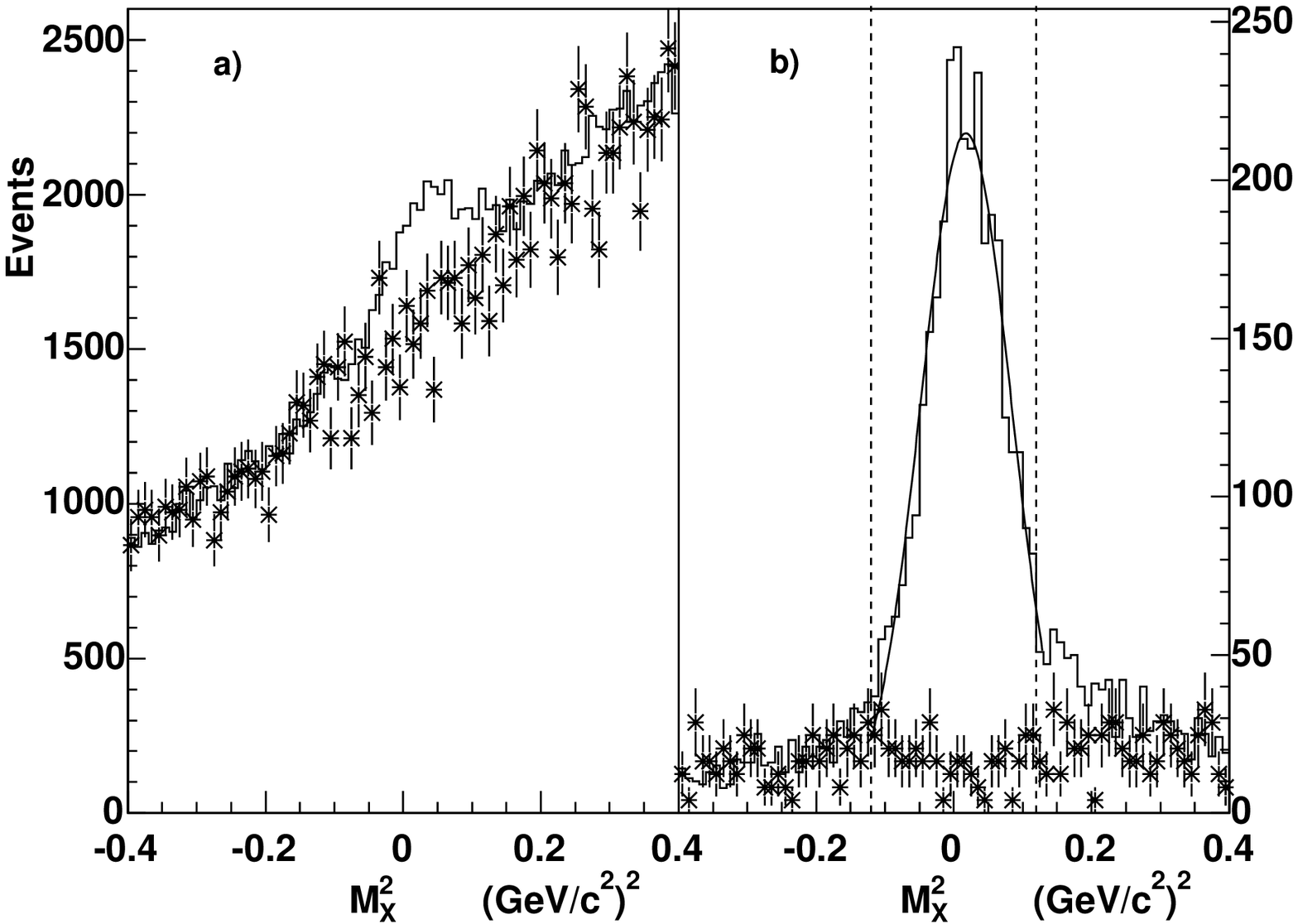}}
\caption{\label{fig:Chen2} 
Missing mass squared $M_x^2$ distributions of the $(e,e'p)x$ reaction
for $(e,e'p\gamma)$ events on a longitudinally polarized N$\vec{\rm H}_3$ target.
{\bf Left:} Raw distribution; {\bf Right:} Distribution after requiring detected photon within $2^\circ$
of the predicted direction of an exclusive photon from the $(e,e'p)\gamma$ kinematics.  In both
Figs.  the stars are the corresponding distributions from a C target, normalized to the negative
$M_x^2$ region.
}
\end{figure}

\begin{figure}
 \centerline{\includegraphics[width=0.64\textwidth]{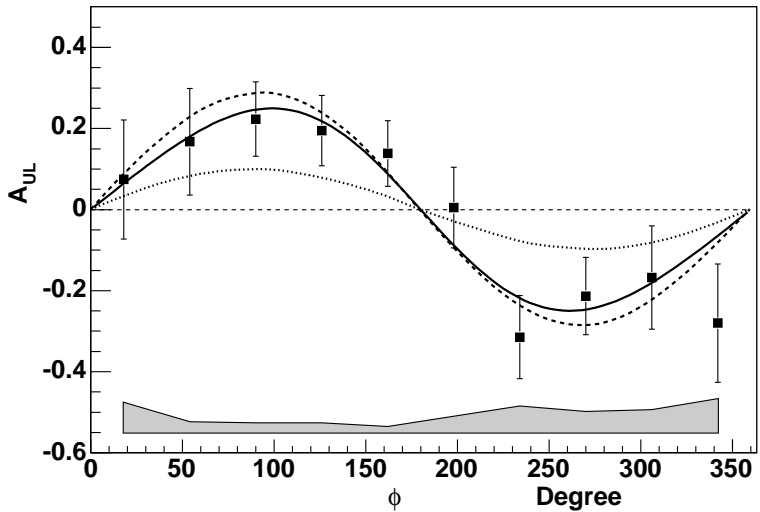}}
 \caption{\label{fig:Chen5} 
 CLAS Longitudinal target spin asymmetry, $A_{UL}$ \cite{Chen:2006na},
 for $\langle Q^2\rangle =1.82 \GeV2$, $\langle\xi\rangle = 0.16$, and
$\langle t\rangle = - 0.31\GeV2$.  The solid curve is a fit of the form of (\ref{eq:BSA-CLAS1}).
The dashed and dotted curves are from the $\xi$-dependent VGG
model with $E = \widetilde{E}=0$.  The dotted curve includes only $H$.  The dashed curve includes both  $H$ and  $\widetilde{H}$.
 }
\end{figure}

\begin{figure} 
\centerline{ \includegraphics[width=0.64\textwidth]{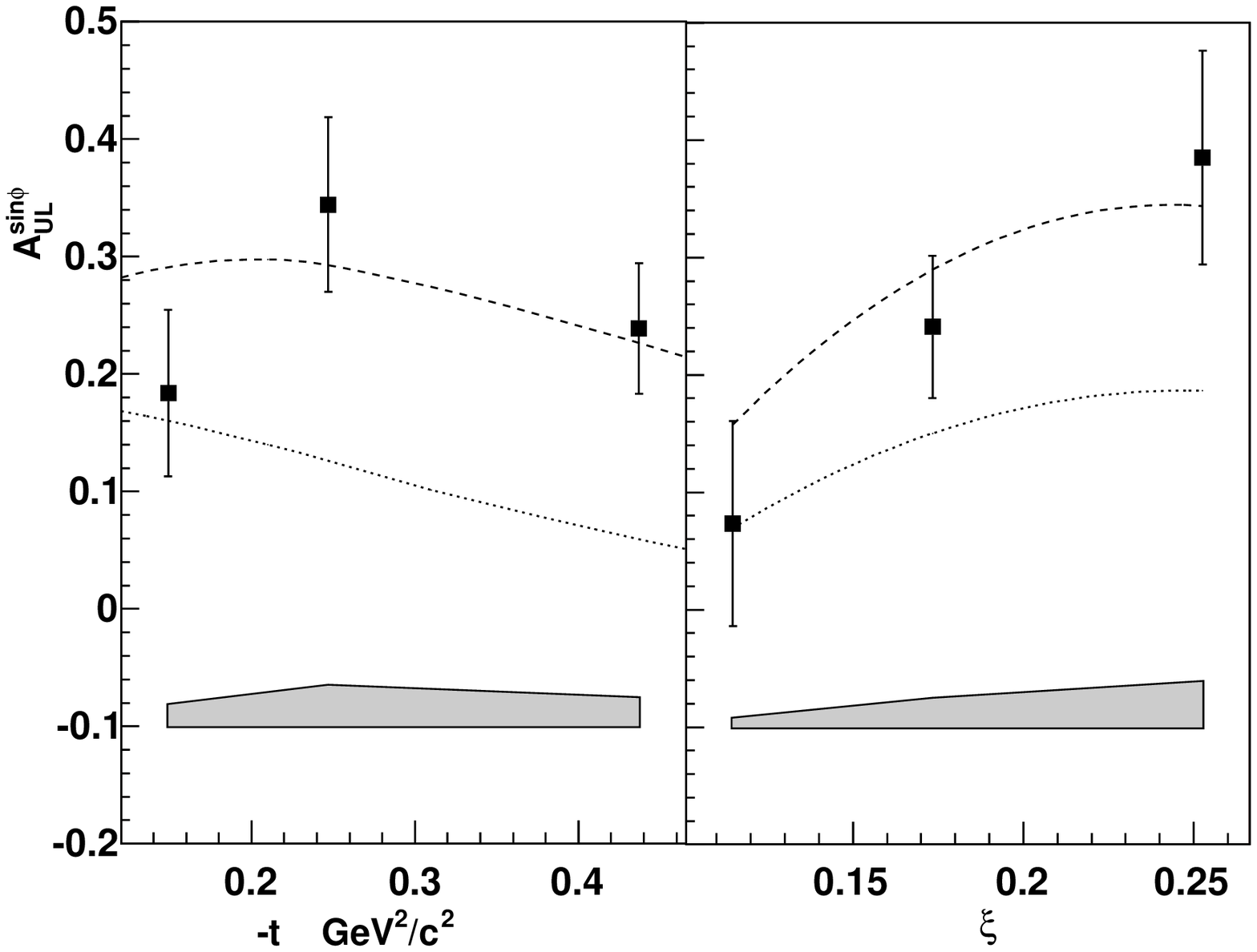}
}
\caption{\label{fig:Chen6} 
The $\sin(\phigg)$ moments of   $A_{UL}$ \cite{Chen:2006na}.
{\bf Left: } Three bins in $-t$, integrated over $\xi$; {\bf Right: }Three  bins in 
$\xi$, integrated over $t$.  The curves are described in Fig.~\ref{fig:Chen5}.
 }
 \end{figure}

A second CLAS experiment, still with the standard CLAS configuration \cite{Mecking:2003zu},
measured the longitudinal target spin asymmetry in the  $\vec{p}(e,e'p\gamma)$ reaction
on a polarized NH$_3$ target
\cite{Chen:2006na}.  In order to isolate the exclusive $ep\rightarrow ep\gamma$ events
from the nuclear continuum, the statistics were limited to the triple coincidence
$(e,e'p\gamma)$ events, with the photons detected in the standard CLAS calorimeter.
The resulting exclusivity spectrum in Fig.~\ref{fig:Chen2} shows a 10:1 signal to background
ratio.
 The longitudinal target spin asymmetry, averaged over the acceptance, is displayed
in Fig.~\ref{fig:Chen5},
 for $\langle Q^2\rangle = 1.82 \GeV2$, $\langle\xi\rangle = 0.16$, and
$\langle t\rangle = - 0.31\GeV2$.  The solid curve is a fit of the same form as (\ref{eq:BSA-CLAS1}).
The resulting $\sin(\phigg)$ moments are plotted in Fig. \ref{fig:Chen6}.
The error bars in Figs.   \ref{fig:Chen5} and  \ref{fig:Chen6} are statistical, with
the systematic errors displayed as a band at the bottom.
The dashed and dotted curves in Figs.   \ref{fig:Chen5} and  \ref{fig:Chen6} indicate
the sensitivity of the longitudinal target spin asymmetry to $\widetilde{H}$.

The initial success of the CLAS DVCS analysis, and the limited small angle acceptance
of the CLAS calorimeter led to the construction of a small angle "Inner Calorimeter".
The ongoing dedicated DVCS program in CLAS will be described in section \ref{sec:CLAS-DVCS}.

\section{The Hall A DVCS Program at 6 GeV}

The Hall A DVCS program started with experiments  E00-110\cite{E00110} 
and E03-106\cite{E03106}. 
These experiments measured, respectively, the cross sections of the
H$(\vec{e},e'\gamma)p$ and D$(\vec{e},e'\gamma)pn$ reactions 
at $\xBj = 0.36$ with an incident beam of 5.75 GeV.
In both experiments the scattered electron was detected in  the standard 
 High Resolution Spectrometer (HRS) \cite{Alcorn:2004sb},  and the
photon was detected in  a new 132 element PbF$_2$ calorimeter,
subtending $\sim 0.1$ sr.     PbF$_2$ is a pure Cerenkov medium, thereby minimizing
the hadronic background and delivering the fastest timing pulses. 
All PbF$_2$ channels were readout by a custom 1 GHz digitizer\cite{Camsonne:2005th}, based
on the ANTARES ARS0 chip\cite{ARS}.
The luminosity of 
1--4 $\cdot 10^{37}\,{\rm Hz}/{\rm cm}^2$ per nucleon
 was  unprecedented  for open
 detectors in a non-magnetic environment.  The halo-free CW beam of CEBAF
 was essential to this success.

\subsection{Proton DVCS}

\begin{figure}
\centerline{\includegraphics[width=0.75\linewidth]{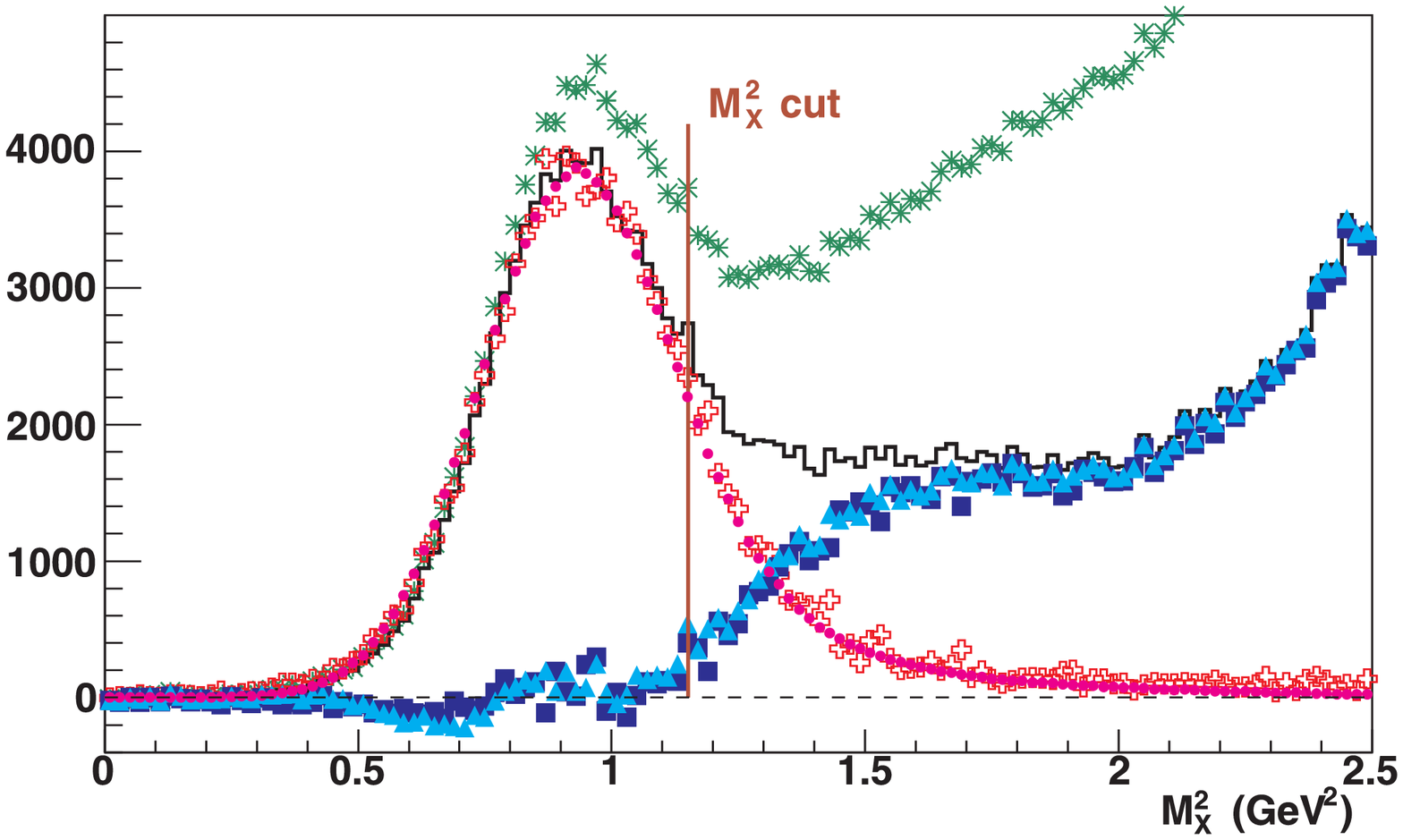}}
\caption{\label{fig:E00110MX2}
Missing mass squared distribution of the H$(e,e'\gamma)X$ reaction in JLab Hall A
experiment E00-110
\cite{MunozCamacho:2006hx}.  
The [green] stars are the raw data after
accidental subtraction.  The continuous [black] histogram is the data after subtracting
the statistical sample of H$(e;e'\gamma)\gamma X'$ events inferred from the measured
H$(e,e'\pi^0)X'$ sample. The open [red] cross histogram is a normalized sample 
of H$(e,e'\gamma p)$ events.  The [magenta] dots are the exclusive simulation.
The [blue] triangles and squares are obtained by subtracting the last two histograms from
the solid [black] histogram.}
\end{figure}

\begin{figure}
\centerline{\includegraphics[width=0.8\linewidth]{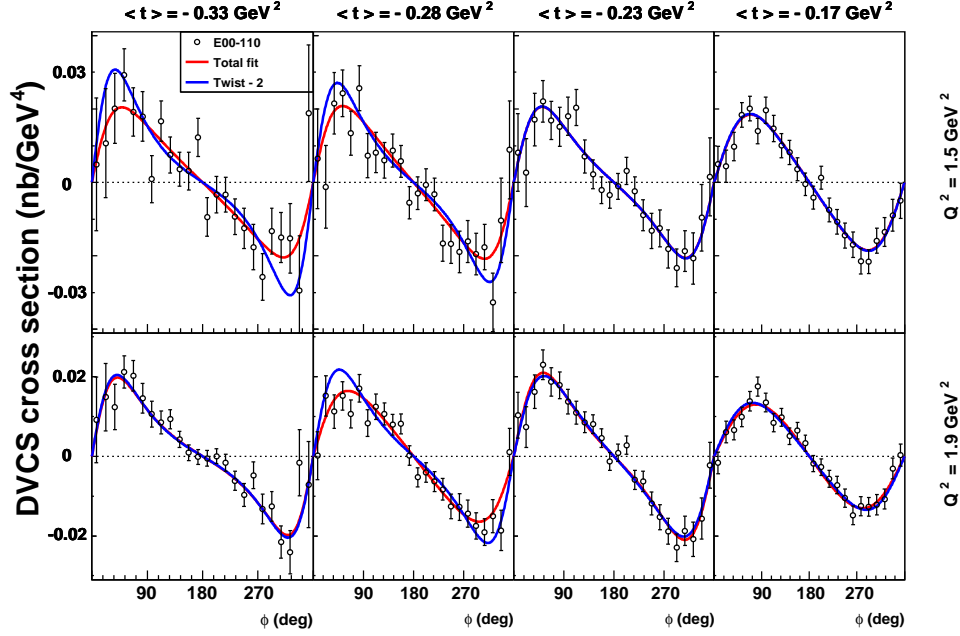}}
\caption{\label{fig:E00110Kin12}
Helicity dependent DVCS cross sections $\Delta^4\sigma/[dQ^2 d\xBj d\Delta^2 d\phiGG]$ from JLab Hall A E00-110 \cite{MunozCamacho:2006hx}
at $Q^2 = 1.5$, and $1.9\,\GeV2$.  The mean values of $-\Delta^2$ are, from right to left
0.17, 0.23, 0.28, and $0.33\,\GeV2$. 
Each distribution is fitted with the form of (\ref{eq:sigmaI}), with the (effective) twist-2 term in red and the
complete fit in blue.
}
\end{figure}

\begin{figure}
\centerline{\includegraphics[width=0.8\linewidth]{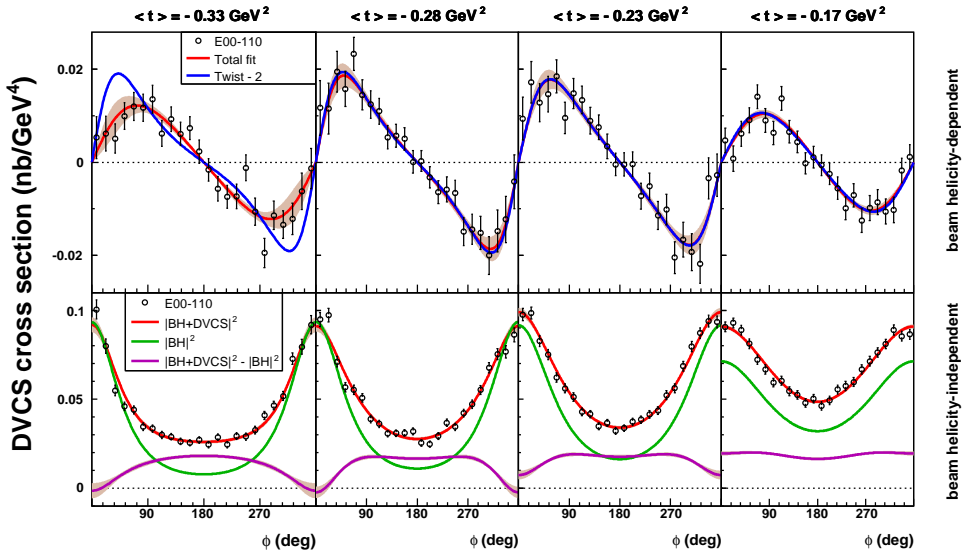}}
\caption{\label{fig:E00110Kin3}
Helicity dependent (top) and independent (bottom) DVCS cross sections from
JLab Hall A at $Q^2 = 2.3\,\GeV2$
\cite{MunozCamacho:2006hx}.
The bins are the same as Fig.~\ref{fig:E00110Kin12}, as are the curves in the top plot.
In the bottom plot, the green curve (mostly concave up) is the pure $|{\rm BH}|^2$ cross section.
The magenta curve (mostly concave down) is a fit including the twist-2 and twist-3
terms of (\ref{eq:sigmaDVCS}, \ref{eq:sigmaI}).
}
\end{figure}

\begin{figure}
\centerline{\includegraphics[width=0.8\linewidth]{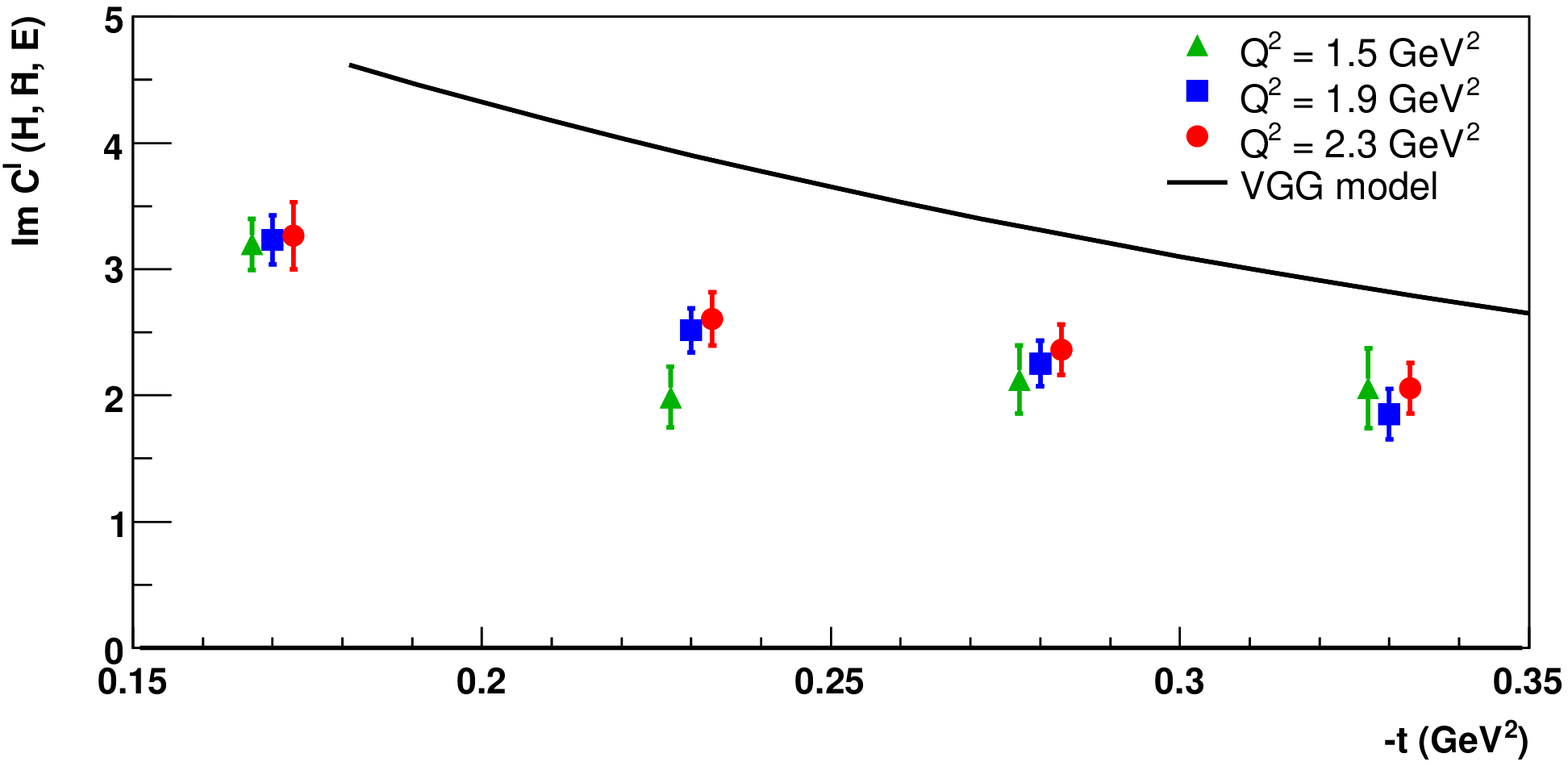}}
\caption{\label{fig:E00110-ImC}
Imaginary part of the effective interference term $C_{\rm unp}^{\mathcal I}$ extracted
from the helicity dependent data of Figs.  \ref{fig:E00110Kin12} and \ref{fig:E00110Kin3}.
}
\end{figure}

Hall A experiment E00-110 measured DVCS on the proton at $Q^2 = 1.5$, 1.9, and $2.3\,\GeV2$.
The isolation of the exclusive H$(\vec{e},e'\gamma)p$ signal is illustrated
in Fig. \ref{fig:E00110MX2}.
The helicity dependent cross sections as a function of $\phiGG$ in four bins in 
$\Delta^2$ are displayed in Figs.  \ref{fig:E00110Kin12} and \ref{fig:E00110Kin3}.
The latter figure also displays the helicity independent cross sections for 
$Q^2 = 2.3\,\GeV2$.  The helicity dependent cross sections demonstrate the
dominance of the effective twist-2 term $s_1^{\mathcal I}$ of (\ref{eq:sigmaI}).  The helicity
independent cross sections (Fig.~\ref{fig:E00110Kin3}) show significant contributions from the sum of
the interference and DVCS terms, in addition to   the pure BH cross section.
Thus the analysis of relative asymmetries of the form $\Delta\sigma/\sigma$ requires
the inclusion of the full DVCS terms in both the numerator and  denominator. 
The effective ``twist-2'' interference term $\Imag {\mathcal C}^I$ of (\ref{eq:CI-unp}) is presented in Fig. \ref{fig:E00110-ImC}.  The VGG model calculation, described in section \ref{sec:VGG}
agrees in slope with the data, but lies roughly $30\%$ above the data.
Within statistics, the results in Fig.  \ref{fig:E00110-ImC} are close to 
$Q^2$-independent in all bins in $\Delta^2$.  This provides 
 support to the
conjecture that DVCS factorization results in leading twist dominance at the same scale
of $Q^2 \ge 2\,{\rm GeV}^2$ as in DIS.

\subsection{Neutron DVCS}

\begin{figure}
\centerline{\includegraphics[width=0.5\linewidth,clip=true]{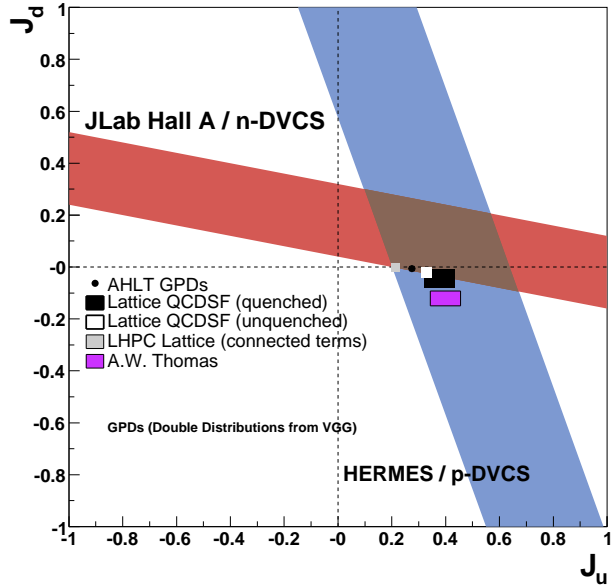}}
\caption{\label{fig:JuJd}
Experimental constraints on the total up and down quark contributions to the proton spin.
JLab Hall A neutron   \cite{Mazouz:2007vj} and HERMES transversely polarized proton 
\cite{:2008jga}.
The theory/model values are from 
AHLT \cite{Ahmad:2006gn}, 
QCDSF quenched \cite{Gockeler:2003jfa} and unquenched \cite{Brommel:2007sb}
LHPC \cite{Hagler:2007xi}, and
Thomas \cite{Thomas:2008ga}.
}
\end{figure}

JLab Hall A experiment E03-106 measured the helicity dependent DVCS cross section 
on deuterium, 
D$(\vec{e},e'\gamma)X$ at $Q^2=1.9\GeV2$ and $\xBj=0.36$.
Within the impulse approximation, the cross section is described as the incoherent sum
of coherent deuteron and quasi free proton and neutron channels:
\begin{equation}
{\rm D}(\vec{e},e'\gamma) = d(\vec{e},e'\gamma)d+n(\vec{e},e'\gamma)n+p(\vec{e},e'\gamma)p+...
\end{equation}
Meson production channels contribute as background.  The proton-DVCS contribution is 
calculated by smearing the H$(e,e'\gamma)X$ data by the nucleon momentum distribution
in the deuteron.  This statistical estimate of the proton contribution is subtracted from the data.  The coherent deuteron and quasi free neutron channels were separated, within statistics, by fitting the
missing mass distribution with a Monte Carlo simulation of these two channels.  This
separation exploits the fact that for $M_X^2$ calculated relative to a nucleon target,  the 
quasi free neutron spectrum peaks at $M_X^2\approx M^2$ whereas the coherent deuteron
peak lies at $M^2+t/2$.
This analysis produced constraints on the neutron and deuteron DVCS$^\dagger$BH 
interference terms $\Imag[C_{\rm unp}^{\mathcal I}]$
\cite{Mazouz:2007vj}.  Mazouz {\em et al.,} \cite{Mazouz:2007vj} fitted the neutron interference
signal by varying the  parameters of the 
$E$ GPD within the VGG model of \cite{Goeke:2001tz}.
This results in a model dependent constraint on the Ji sum rule values of $(J_d,J_u)$, illustrated
in Fig.~\ref{fig:JuJd}.  A similar constraint obtained by the HERMES collaboration 
in DVCS on a transversely polarized proton target is also illustrated in the figure.  Both of
these experimental determinations are essentially constraints on the model at one value of $\xBj$,
and then the model is integrated over $x$ at fixed $\xi$ to obtain the sum rule estimate.
Measurements over a more extensive range in $\xi$ with a more complete set of 
spin observables, and models with more degrees of freedom are necessary in order to
more fully constrain the sum rule with realistic error bars.  Lattice QCD calculations, and other phenomenological 
estimates are also illustrated in Fig.  \ref{fig:JuJd}. 

\subsection{Future Hall A Program at 6 GeV}

The unpolarized cross sections in Fig.~\ref{fig:E00110Kin3} are not fully dominated by the pure
BH process.  The harmonic $\phiGG$ structure of the cross section does not allow the
full separation of the $\Real[{\rm DVCS}^\dagger{\rm BH}]$ and $|{\rm DVCS}|^2$ contributions.
These terms can be separated either by the beam charge dependence 
({\em e.g.} \cite{Airapetian:2006zr}) or by measuring the incident energy dependence of the
cross sections.  At fixed $Q^2$, $\xBj$, the DVCS, Interference, and BH terms in the 
cross section scale roughly as $s_e^2 : s_e : 1$ (\ref{eq:sigmaDVCS}, \ref{eq:sigmaI}).
Experiment E07-007 
\cite{E07-007}
will measure the DVCS helicity independent cross 
sections in the three kinematics of Figs.~\ref{fig:E00110Kin12} 
and \ref{fig:E00110Kin3}
at two separate beam energies in each kinematics.  This will measure the $Q^2$
dependence of the separated
leading twist-2 and twist-3 observables of the  $\Real[{\rm DVCS}^\dagger{\rm BH}]$
 and $|{\rm DVCS}|^2$ terms.

Experiment E08-025
\cite{E08-025} will measure the DVCS cross sections on the deuteron at the same 
$(Q^2,\xBj)$ value as in E03-106, but at two incident beam energies.  Together with an expanded calorimeter to improve the neutral pion subtraction, the two beam energies will allow a more
complete separation of the DVCS$^2$ and  real and imaginary parts of the DVCS$\cdot$BH interference
on a quasi-free neutron.  This will be an important step towards a full flavor separation of  DVCS.
Both experiments E07-007 and E08-025 are running in Autumn 2010.

\section{The CLAS DVCS Program at 6 GeV}
\label{sec:CLAS-DVCS}

\subsection{Unpolarized Proton Targets}

A new calorimeter of 424 tapered PbWO$_4$ crystals  was constructed to provide 
complete $2\pi$ photon coverage for polar angles from 4.5$^\circ$ to 
15$^\circ$, relative to the beam line. A $\approx$ 5 Tesla superconducting 
solenoid was  added at the target, to confine Moeller electrons. The 
new calorimeter is located 60 cm from the target
where the solenoid fringe field is still a few Tesla.
Therefore, the individual crystals were read-out by Avalanche Photo-Diodes. 
Having strongly
benefited from the CERN CMS pioneering research and development effort on this
recent technology, the present CLAS experiment is the first one to use 
such photodetectors in a physics production mode.

\begin{figure}
\centerline{\includegraphics[width=0.8\linewidth]{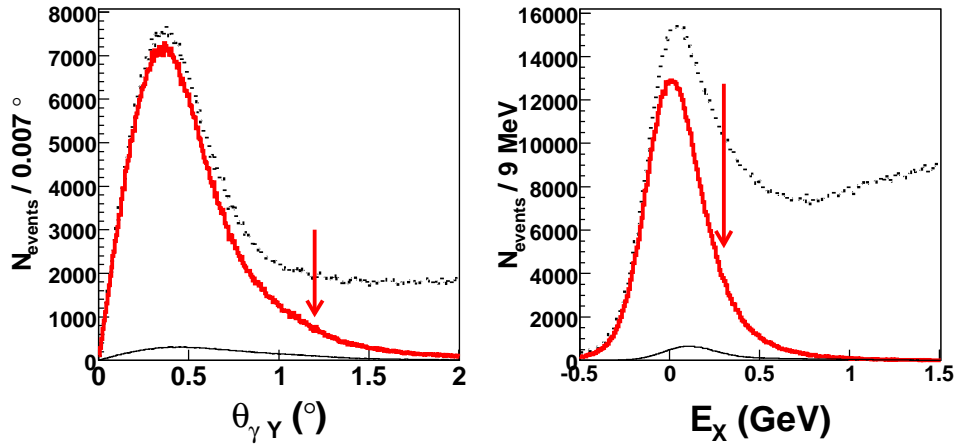}}
\caption{\label{fig:CLAS-Exc}
Example of exclusivity cuts (given by the location of the arrow) for 
the CLAS DVCS experiment\cite{Girod:2007jq}.
Distribution in cone angle $\theta_{\gamma Y}$ for the $ep\to epY$
reaction (left) and in missing energy $E_X$ for the $ep\to ep\gamma X$
reaction (right) before (black-dotted curve) and after (red solid)
kinematic cuts (including others not displayed here). The thin solid black line
represents the background from the $ep\to ep\pi^0$ events. These distributions are
integrated over all kinematics variables.}
\end{figure}

All particles of the $ep\to ep\gamma$ reaction final state
were detected in CLAS. To ensure exclusivity, several cuts were made, a couple
of them being illustrated in Fig.\ref{fig:CLAS-Exc}. In spite 
of these very constraining cuts, some contamination from the $ep\to ep\pi^0$ reaction 
remained. Indeed, when one of the two $\gamma$'s originating from the decay 
$\pi^0\hookrightarrow\gamma \gamma$ escapes detection
and/or has little energy (below the 150 MeV threshold of the calorimeter), an 
event $e p \rightarrow ep \gamma(\gamma)$ may pass all
DVCS cuts and become a perfect candidate
to be selected as an $ep\to ep\gamma$ event. Such ``1-$\gamma$" $\pi^0$ background 
can be estimated from Monte-Carlo combined with the actual number of
detected ``2-$\gamma$" $\pi^0$'s, resulting, depending on the kinematics, in 
contaminations ranging from 1 to 25\%, being 5\% in average.

\begin{figure}
\centerline{\includegraphics[width=0.8\linewidth]{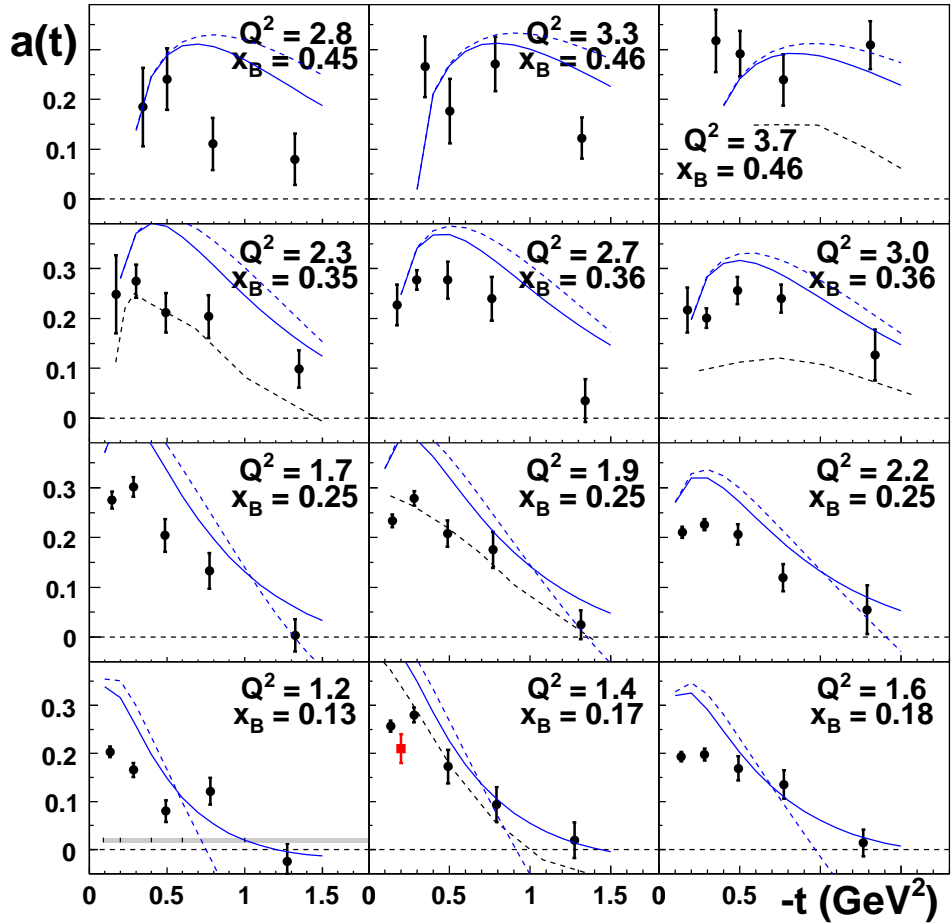}}
\caption{\label{fig:FXfig4}
Beam spin asymmetry $\sin(\phigg)$ moments from the CLAS DVCS experiment
\cite{Girod:2007jq}. Curves are described in the text.}
\end{figure}

The extensive CLAS 
dataset of DVCS beam spin 
asymmetries (BSA) is illustrated in Fig.~\ref{fig:FXfig4}. The blue solid curves are 
the result of the twist-2 handbag GPD calculation (VGG) including just the $H$ GPD 
\cite{Vanderhaeghen:1999xj,Guidal:2004nd}.
The blue dashed curves include the associated 
twist-3 calculation.  Although the  general trends
of the data are reproduced, 
the
model  tends to overestimate 
the BSAs. These too large BSAs by the VGG model can come from either an 
overestimation of $\mathcal{H}(\xi,t)$ (the dominant factor in the numerator of the BSA)
or an underestimation of the CFFs associated to the real part of the DVCS amplitude,
which contribute predominantly to the denominator of the BSA
\cite{Guidal:2008ie}.
The dashed black curve (third curve in some panels of Fig.\ref{fig:FXfig4})
is the result 
of a Regge  model~\cite{Laget:2007qm} 
for the DVCS process.
As $Q^2$ increases, the Regge model drops significantly below both the data and the VGG
calculations.
This experiment was continued in  2008--2009, which will significantly improve
the statistical precision, relative to Fig.\ref{fig:FXfig4} \cite{E06-003}.

\subsection{Polarized Targets}

In  2009, a new DVCS experiment completed data taking with the longitudinally 
polarized NH$_3$ target
\cite{E05-114}.  
Relative to the previous experiment 
(\cite{Chen:2006na} and Figs.~\ref{fig:Chen2}--\ref{fig:Chen6}), this new experiment 
will improve both the statistics and acceptance by the addition of the new 
electromagnetic calorimeter mentioned in the previous section. We recall that
the target spin asymmetry is mostly sensitive to $\widetilde{\mathcal{H}}(\xi,t)$ and that,
therefore, strong constraints on this CFF should arise from this experiment, as discussed
{\em e.g.} in \cite{Guidal:2010ig}.

After decades of development, the HD-ice target ran successfully at the BNL-LEGS
facility in 2005 and 2006.  
This target has now been transferred to JLab and is being
prepared for a photo-production run in 2011 \cite{E06-101}.  Initial studies of local depolarization by
microwaves suggest that the spin relaxation times of this target are sufficiently long
for the target to operate with electron beams in CLAS.  An electron beam test is projected 
for the end of the 2011 photo-production run.  If successful, a full suite of transverse
polarization observables for the DVCS process will be feasible
in CLAS in 2011.  When combined with the cross section and longitudinal polarization 
data, along with the resulting double polarization observables,
a full separation of the real and imaginary parts of all four Compton Form Factors
$\mathcal{H}(\xi,t)$, $\mathcal{E}(\xi,t)$, $\widetilde{\mathcal{H}}(\xi,t)$, and
$\widetilde{\mathcal{E}}(\xi,t)$ is in principle possible~\cite{Belitsky:2001ns,Guidal:2008ie}.

\subsection{Nuclear Targets}

GPDs are also  defined for nuclei \cite{Berger:2001zb,Cano:2003ju}.  One can study effects similar to 
the EMC effect observed for standard inclusive parton distributions functions (PDF)
where the PDF of a nucleus is not simply the sum of the individual nucleon PDFs.
A pioneering experiment~\cite{E08-024,Voutier:2008wu} of  coherent
DVCS on a $^4$He target
ran with the CLAS detector in 2010.   
$^4$He is a very good 
starting case study as it is dense enough to generate nuclear medium effects, many 
microscopic calculations for its nuclear structure and dynamics exist and,
as a global spin-0 object, at leading-twist, there is only one GPD.
This $^4$He-DVCS experiment detected the scattered 
electron in CLAS, the final state photon with the PbWO$_4$ and standard  calorimeters 
mentioned in the previous sections and the recoil nucleus with a radial time-projection 
chamber~\cite{Fenker:2008zz}.
The $\phi$-distribution of the coherent DVCS BSA, up to twist-3 corrections,  can yield
the real and imaginary parts of the Compton Form Factor of the coherent  GPD.

\section{Deeply Virtual Meson Production}
\label{sec:VectorMeson}

GPDs are in principle also accessible through exclusive 
meson electroproduction  (see Fig.~\ref{fig:DVCS-GPD}).
With respect to the DVCS process, a few features are proper to 
the meson channels:
\begin{itemize}
\item The factorization holds only for the longitudinal part 
of the amplitude which implies to separate, experimentally, the
transverse and longitudinal parts of the cross section.
For pseudo-scalar mesons, this can be done through a Rosenbluth 
separation.     
For vector mesons, this separation can be carried out, relying on the s-channel 
helicity conservation (SCHC) concept, by measuring the
angular distribution of the vector meson decay products.
\item In comparison to the DVCS handbag diagram, there is now a perturbative gluon
exchange.   This suggests that factorization will be obtained at a 
higher scale in exclusive meson production than in DVCS.
\item Besides the GPDs, there is another non-perturbative
object entering the meson handbag diagram,  the meson distribution
amplitude (DA). It is usually taken as the asymptotic DA 
but it  potentially adds a further unknown in the process.
\item As a positive point,  the meson channels have the advantage of filtering
certain GPDs: the vector meson channels are sensitive, at leading twist, only
to the $H$ and $E$ GPDs while the pseudoscalar mesons are sensitive only to
the $\tilde{H}$ and $\tilde{E}$ GPDs. 
Deeply virtual meson production also offers a flavor filter of the GPDs.  For example,
$\rho^0$ and $\omega$ electroproduction are sensitive to different combinations of
the up- and down-quark GPDs.
\end{itemize}
Exclusive $\pi^0$ electroproduction results have been published from CLAS \cite{DeMasi:2007id} and Hall A \cite{Fuchey:2010}.
The Hall C results on exclusive $\pi^+$ production are discussed in the ``Transition to Perturbative QCD'' chapter of this volume
\cite{Gilman:2010}.
In this section, we focus on exclusive vector meson production.

\subsection{The $\rho^0$ channel}

Deeply virtual electroproduction of the $\rho^0$ was studied by the CLAS collaboration
at incident energies of 4.2 GeV
\cite{Hadjidakis:2004zm}
 and  5.75 GeV
\cite{Morrow:2008ek}.
To select the  $e^-p \rightarrow e^-p\rho^0\hookrightarrow \pi^+\pi^-$ channel,  
the scattered electron, the recoil proton and the  $\pi^+$
were detected. A cut on the missing mass $ep\to ep\pi^+X$ was then used 
to  identify the $e^-p \rightarrow e^-p\pi^+\pi^-$ 
final state.
The main challenge in this analysis was to subtract under the broad
($\Gamma_{\rho^0}\approx 150$ MeV) $\rho^0$ peak, the non-resonant 
$e^-p \rightarrow e^-p\pi^+\pi^-$ (physical) background, arising for 
instance from processes such as 
$e^-p \rightarrow e^-\pi^-\Delta^{++} \hookrightarrow p\pi^+$.
These  ``background'' channels led to uncertainties of the order 
of 20\% to 25\% on the extracted cross sections. 

\begin{figure}
\centerline{\includegraphics[width=0.95\linewidth]{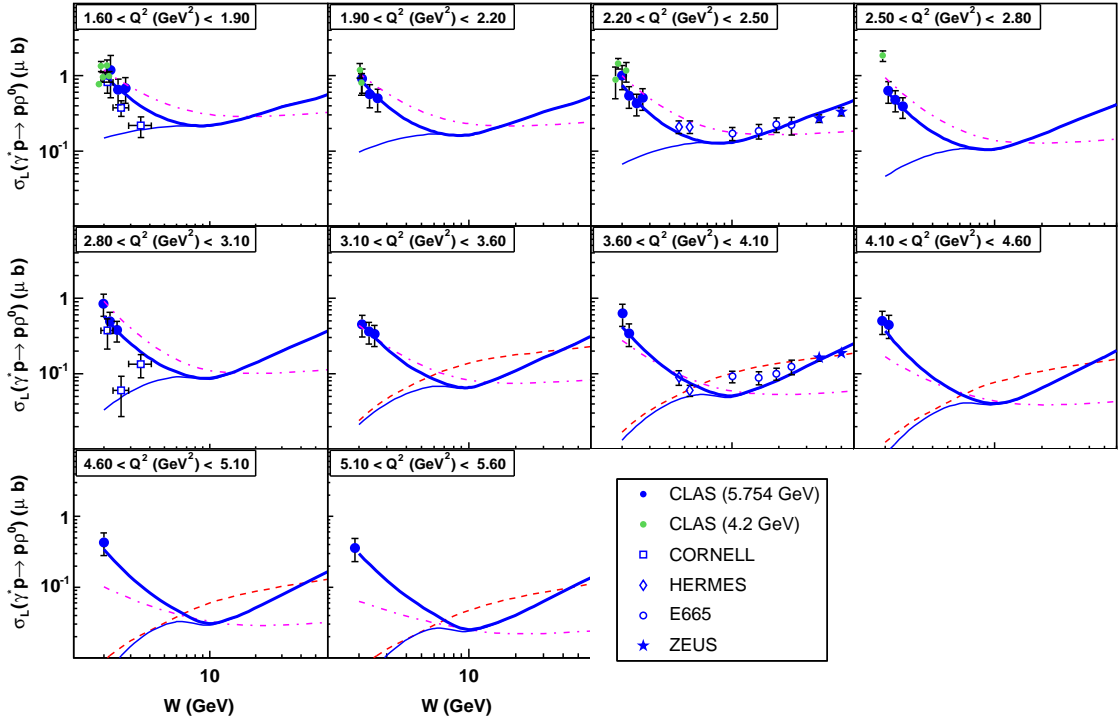}}
\caption{\label{fig:rhow}
World data (for  $W<50$ GeV) for the reduced cross sections 
	$\gamma^*_L p \rightarrow p \rho^0_L$ 
	as a function of $W$ for constant $Q^2$ bins
	($\mu$b). The dashed curve  
	(GK) \cite{Goloskokov:2006hr,Goloskokov:2005sd} 
	 and the thin solid curve  
	(VGG) \cite{Vanderhaeghen:1999xj} 
are GPD calculations. The thick solid curve is the
	VGG calculation with the addition of the D-term inspired contribution 
	\cite{Morrow:2008ek,Guidal:2007cw}. 
	The dot-dashed  curve is the Regge JML~\cite{Laget:2000gj} calculation.
	The 5.75 GeV CLAS, 4.2 GeV CLAS, CORNELL,
 HERMES, FermiLab and ZEUS data are 
	respectively from refs.~\cite{Morrow:2008ek}, \cite{Hadjidakis:2004zm},
	\cite{Cassel:1981sx}, \cite{Airapetian:2000ni},
	\cite{Adams:1997bh},	
	 and \cite{Chekanov:2007zr}.
}
\end{figure}

The separation of the longitudinal and 
transverse parts of the cross section was carried out, as mentioned earlier,
by studying the angular distribution of the decay pions in the center of mass of the
$\pi^+\pi^-$ system.  At the same time,  by the analysis of various azimuthal 
angular distributions, SCHC was verified experimentally  at the $\approx$ 20\% level. 
The 
longitudinal part of the 
$\gamma^* p \rightarrow p\rho^0$ cross section, which can in principle lend 
itself to a GPD interpretation, is displayed in Fig.~\ref{fig:rhow} along with
the world data.

The cross sections clearly exhibit two different behaviors as
a function of $W$. At low $W$, the cross sections decrease 
as $W$ increases ($\xBj $ decreases)  and then begin to rise slowly for $W> 10$  GeV.
These two kinematic regimes can be identified, 
 simply speaking,  with regimes of $t$-channel exchange  of 
Reggeon or $q\bar{q}$ exchange in the former case  and of
Pomeron or 2-gluon exchange in the latter case.
The results of the calculations of the JML~\cite{Laget:2000gj} model, 
based on Reggeon exchange and hadronic degrees of freedom, 
and of the VGG~\cite{Vanderhaeghen:1999xj} and GK~\cite{Goloskokov:2006hr,Goloskokov:2005sd} 
models based on GPDs and on the 
handbag diagram of Fig.~\ref{fig:DVCS-GPD} are shown in Fig.\ref{fig:rhow}. At lower $W$ values, 
where the new CLAS data lie, it is striking that both the GK and VGG models 
fail to reproduce the data even though they are very successful at large $W$, even at 
$Q^2\approx 2.3\GeV2$. In the high-$W$ (low $\xBj$) region, the gluon GPD calculations
already contain large higher-twist effects in the form of intrinsic $k_\perp$ effects.
The question then arises whether the higher twist effects 
have a different nature
in the region dominated by quark GPDs  (low $W$), or whether the 
double distribution based GPD models are missing an
essential contribution.
Ideas for such ``missing" contribution in the $D$-term of the  GPDs 
are speculated 
in  \cite{Morrow:2008ek,Guidal:2007cw}, leading to the thick solid curve in Fig.
\ref{fig:rhow}.

\subsection{The $\omega$ channel}

The $\omega$ channel was studied in CLAS
 by detecting the $e^-p \rightarrow e^-p\pi^+X$
and $e^-p \rightarrow e^-p\pi^+\pi^-X$ topologies
 \cite{Morand:2005ex}.
The former is advantageous to
determine total cross sections with high statistics and the latter is necessary to measure the
distribution of the decay products of the $\omega$  to separate
the longitudinal and transverse parts of the cross section if SCHC is verified.
However, one important result of this experiment was that many
SCHC-violating spin density matrix elements were measured to be
significantly different from 0 in $\omega$ electroproduction. Therefore, the longitudinal and
transverse parts of the cross sections were not separated.
Also, the angular analysis revealed the importance of
unnatural parity exchange in the $t$-channel, such as $\pi^0$ exchange. 

\begin{figure}
\includegraphics[width=0.45\textwidth]{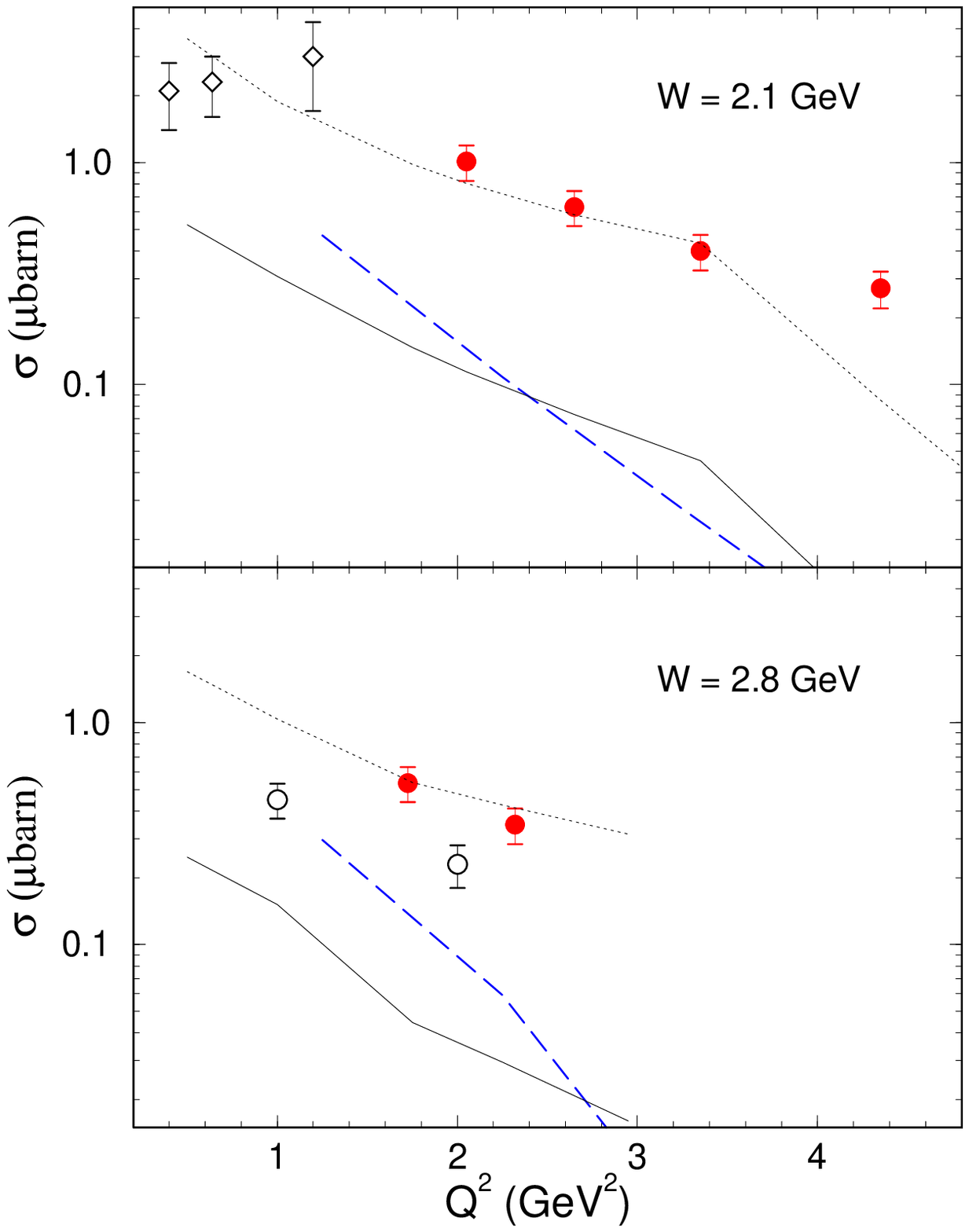}
 \includegraphics[width=0.45\textwidth]{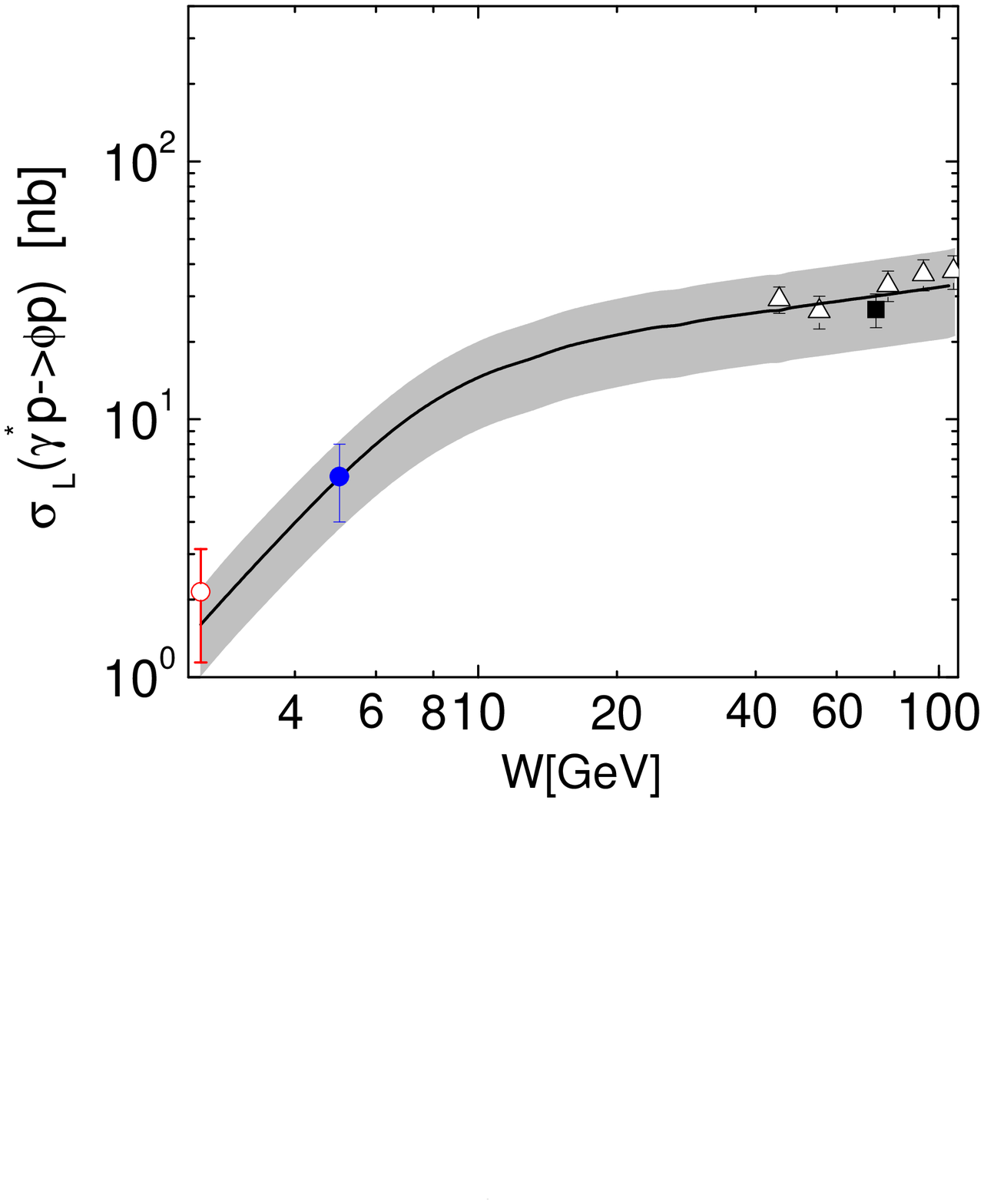}
\caption{\label{fig:omega}\label{fig:phi}
{\bf Left:} Total (unseparated) cross section for the reaction 
$\gamma^*p \rightarrow p\omega$, for 
\mbox{$\langle W \rangle $=2.1 GeV} (top) and $\langle W \rangle $=2.8 GeV (bottom),
as functions of $Q^2$. The dotted curve is 
the JML model for the total cross section $\sigma_T+\epsilon\sigma_L$
and the solid (dashed) curves shows the result of the JML (VGG) calculation for
$\epsilon\sigma_L$. Data are from CLAS (full circles)~\cite{Morand:2005ex}, 
DESY (empty diamonds)~\cite{Joos:1977tz} and Cornell (empty circles)~\cite{Cassel:1981sx}.
{\bf Right:} Longitudinal 
$\gamma^*p \rightarrow p\phi$ cross section as a function of W at $Q^2$=3.8 GeV$^2$.
Data from CLAS (open circle)~\cite{Santoro:2008ai}, HERMES (solid circle)~\cite{Borissov:2001fq}, 
ZEUS (open triangles)~\cite{Chekanov:2005cqa} and H1 (solid square)~\cite{Adloff:2000nx}.
(Courtesy of P. Kroll and S. Goloskokov).}
\end{figure}

In terms of quantum numbers, the $\pi^0$ exchange contribution can be identified to the 
$\tilde{E}$ GPD. In the framework of the JML model \cite{Laget:2004qu},  
$t$-channel $\pi^0$ exchange is a major contributor to  the cross section.
The suggested importance of $\pi^0$-exchange is in apparent  contradiction with the theoretical prediction
that,  
at sufficiently large $Q^2$, exclusive vector meson production 
should be mostly longitudinal and sensitive only to $H$ and $E$.
Therefore, in order to study the GPD formalism in $\omega$ production,
it is essential to experimentally isolate  the purely longitudinal cross section, via
a Rosenbluth separation.
The VGG calculation of $\sigma_L$, shown in Fig.\ref{fig:omega}, 
lies well below the unseparated data.   This suggests that a precision extraction
of $\sigma_L$ via a Rosenbluth separation will be a difficult experimental challenge.

\subsection{The $\phi$ channel}

The $e^-p \rightarrow e^-p\phi\hookrightarrow K^+K^-$ reaction was identified in CLAS 
by detecting the scattered electron, the recoil proton and the positive kaon and cutting
around the missing mass of a kaon
~\cite{Santoro:2008ai}. Relying on the SCHC concept, which was
experimentally verified to hold in this channel, the longitudinal/transverse separation of
the cross section was carried out. Fig~\ref{fig:phi} shows the resulting 
longitudinal total cross section at $Q^2=3.8\,\GeV2$, along with  higher energy HERMES and HERA data at comparable $Q^2$.

 Exclusive $\phi$ electroproduction on the proton can be interpreted in terms 
of the handbag diagram with {\it gluonic} GPDs.
Fig~\ref{fig:phi} shows the result of such calculation in the framework of the GK 
model~\cite{Goloskokov:2006hr,Goloskokov:2005sd}. 
The very good agreement between this GPD calculation and
the data gives confidence in the 
way higher twists corrections are handled, {\em i.e.} by taking into account the 
intrinsic transverse momentum dependence of the partons in the handbag
calculation.

This set of three experiments delivered the largest ever dataset  on vector meson production in the
large $Q^2$ valence region.
Although conclusions for the meson channels are more challenging than for DVCS,
there may be the possibility to interpret the $\rho^0$ and $\phi$ channels in terms of the handbag diagram,
though with large higher-twist corrections and possibly  modifications of the
Double-Distribution based GPD parametrisations.  
The higher $Q^2$ data from JLab at 12 GeV, as well as a global analysis including the larger
DVCS dataset anticipated in the coming years should greatly clarify the role of factorization
in deep virtual vector meson production.

\section{Outlook}
\label{sec:Outlook}

\subsection{Jefferson Lab at 12 GeV}

\begin{figure}[htbp]
\includegraphics[width=0.475\linewidth]{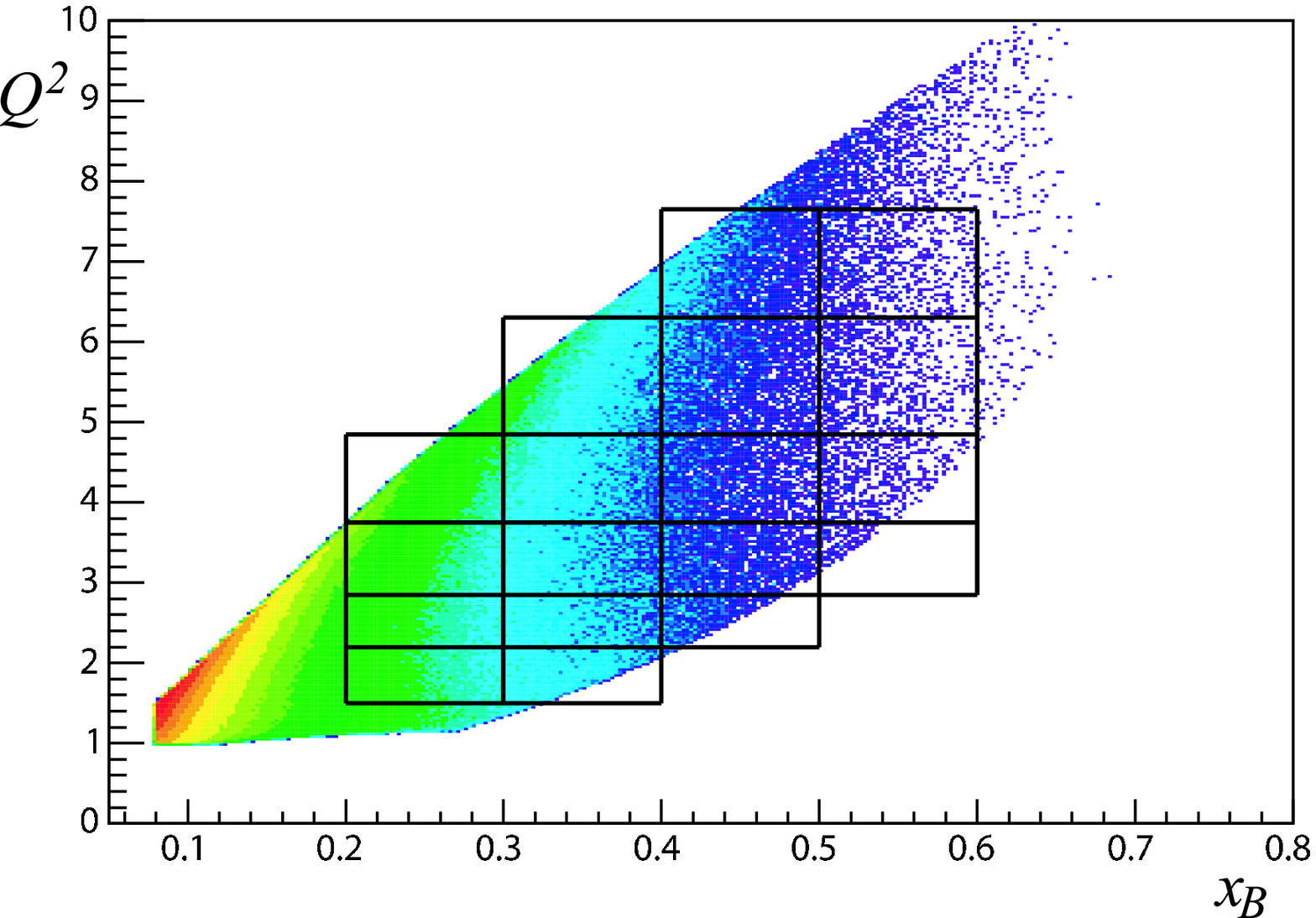}\hfill
\includegraphics[width=0.475\linewidth]{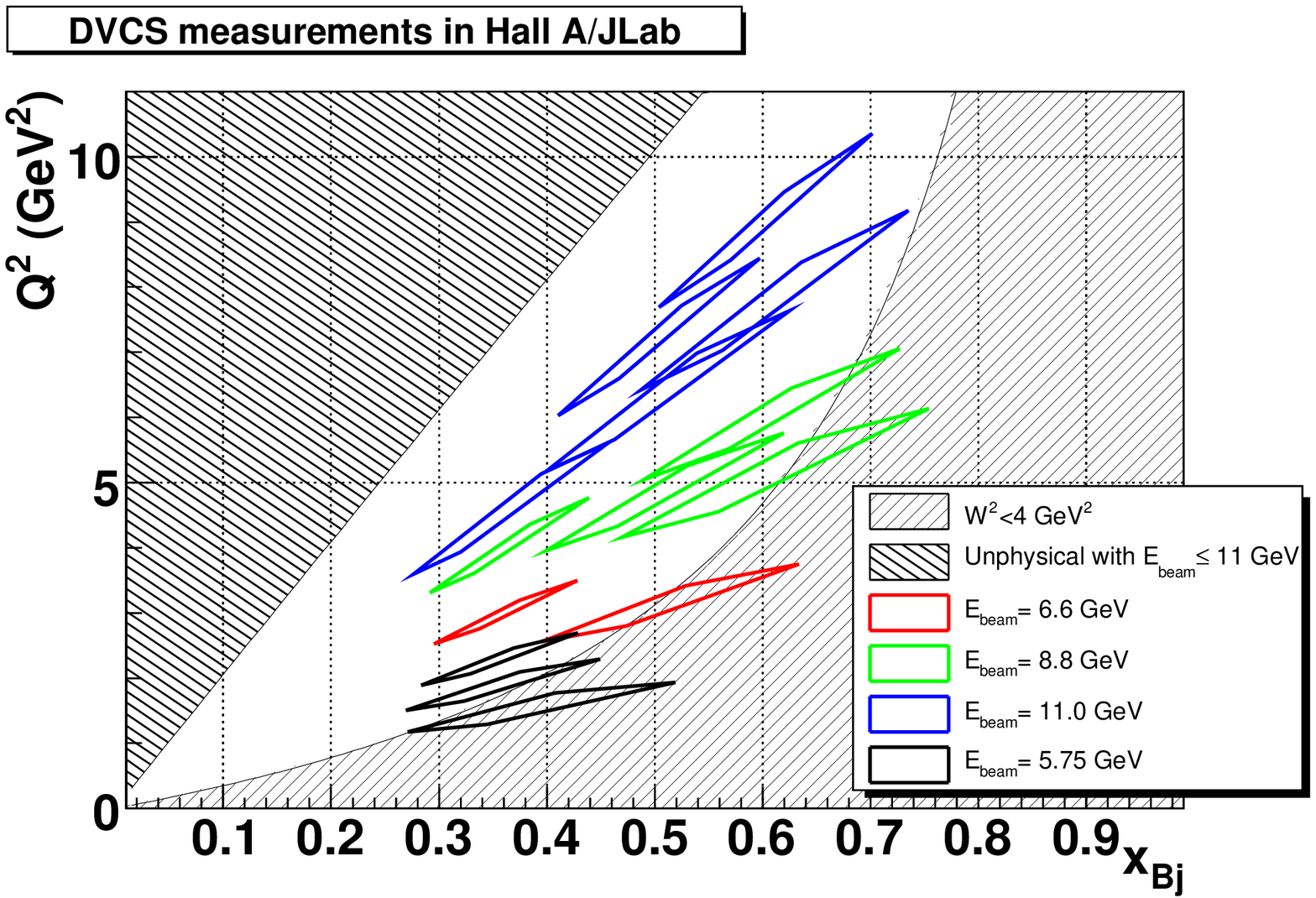}
\caption{\label{fig:DVCS12GeV}
Projected kinematic bins at JLab 12 GeV;
{\bf Left:} CLAS12 kinematics for DVCS and DVMP on unpolarized H$_2$ 
and longitudinally polarized NH$_3$ targets.  The colors and density are proportional
to the relative count rates.
{\bf Right:}  Hall A kinematics for DVCS and $\pi^0$ electroproduction.  Beam time is adjusted for roughly 
equal counts in all bins.
}
\end{figure}

The JLab 12 GeV project offers an unprecedented frontier of intensity
and precision for the study of deep exclusive scattering.
The design luminosity of the upgraded CLAS12 detector is $10^{35}/{\rm s\,cm}^2$,
with a large phase space acceptance for simultaneous detection of DVCS and
deeply virtual meson production channels.  At this luminosity, 
the Hall B dynamic nuclear polarization NH$_3$ target will achieve a longitudinal proton polarization 
of  $80\%$.
The Hall A and Hall C spectrometers will allow dedicated studies at luminosities 
$\ge 10^{37}/{\rm s\,cm}^2$ for neutral channels $\gamma,\, \pi^0$ at low $t$
and up to $4\cdot 10^{38}/{\rm s\,cm}^2$ for charged channels $\pi^\pm,\,K^\pm$.
Specific 12 GeV experiments on hydrogen are approved  in Hall A for DVCS
(E12-06-114), in Hall B
for DVCS (E12-06-119),  and deep virtual $\pi^0,\, \eta$ production (E12-06-108), and in Hall C
for deep virtual $\pi^+$ production (E12-06-101,   E12-07-105).
Detailed descriptions of these experiments are available on the Hall A, B, and C web pages
at {\tt www.jlab.org}.
The projected kinematic range of the DVCS programs in Hall A and B is illustrated in
Fig.\ref{fig:DVCS12GeV}.
With CLAS12, additional studies are in progress for measurements of
deep virtual vector meson production,
neutron DVCS via D$(e,e'\gamma n)p$ (LOI-09-001), coherent deuteron DVCS (PR-06-015)
and DVCS on transversely polarized protons.

\subsection{Beyond 12 GeV}

The COMPASS experiment at CERN proposes to measure DVCS in high energy muon
scattering  at low $\xBj$ via triple coincidence H$(\vec{\mu}^\pm,\mu^\pm,\gamma p)$ detection \cite{d'Hose:2004kj}.
The muon beams have the particularity that the muon spin and charge are correlated, enabling
 measurements of the DVCS$^\dagger\cdot$BH interference via correlated beam charge-spin asymmetries.
 In addition to the COMPASS spectrometer, exclusivity will be determined by detecting the
 recoil protons  in a scintillation   array surrounding the target.
 The expected (correlated) range for DVCS and exclusive vector meson production
  is $\xBj \in(0.03,\,0.25)$ and $Q^2\in(1.5,\, 7.5)\GeV2$. 
 A future electron ion collider, with luminosity several orders of magnitude higher than HERA would greatly expand the
 reach of GPD studies.  Maximizing the luminosity is essential to measure fully differential cross sections in all kinematic
 variables.  A collider can deliver both longitudinally and transversely polarized beams without the accompanying background
 of unpolarized nuclei of polarized targets.  A collider would also offer enhanced opportunities for spectator tagging
 to measure neutron GPDs, and recoil tagging for nuclear GPDs.
 
\subsection{Conclusions} 
Deep virtual exclusive scattering offers the tantalizing prospect of forming
spatial images of quarks and gluons in the nucleon.
The GPD formalism has already given us new insight into
nucleon structure, with evidence for quark angular momentum
emerging from GPD models and lattice calculations,  and global
analysis of forward parton distributions and electromagnetic form
factors.  A very important study of DVCS and DES in the valence
region has started with JLab at 6 GeV and will expand with the
12 GeV upgrade.  Several systematic analysis demonstrate the important
constraints on individual GPDs  of the proton and neutron obtained from the data
\cite{Kumericki:2009uq,Guidal:2010ig,Moutarde:2009fg}.
The unprecedented quality of the CEBAF continuous wave beam is 
essential to  achieving full exclusivity at high luminosity.    The revolution in polarized
 beams and targets over the past two decades allows us a full study
of the spin degrees of freedom of DES.
Over the next decade, JLab and COMPASS will obtain new precision DVCS data spanning 
 a factor of 20 in $\xBj$, and at each value of $\xBj$, a factor of two in $Q^2$, with maximal
 $Q^2$ from 4 to 10 $\GeV2$.  The present JLab data are fully differential in $Q^2$, $\xBj$
 and $t$, allowing a systematic study of the approach to scaling in both cross section and
 asymmetry observables.

\ack
This work was supported by US DOE and French CNRS/IN2P3 and ANR. The authors thank 
our colleagues whose spirited conversations have sharpened and deepened our understanding
of this subject. 

 This paper is authored by Jefferson Science Associates,
LLC under U.S. DOE Contract No. DE-AC05-06OR23177. 

\section*{References}
\bibliography{jlab_chapter6}

\providecommand{\newblock}{}
\begin{thebibliography}{100}
\expandafter\ifx\csname url\endcsname\relax
  \def\url#1{{\tt #1}}\fi
\expandafter\ifx\csname urlprefix\endcsname\relax\def\urlprefix{URL }\fi
\providecommand{\eprint}[2][]{\url{#2}}

\bibitem{Ji:1996ek}
Ji X~D 1997 {\em Phys. Rev. Lett.\/} {\bf 78} 610--613 (\textit{Preprint}
  \eprint{hep-ph/9603249})

\bibitem{Radyushkin:1996nd}
Radyushkin A~V 1996 {\em Phys. Lett.\/} {\bf B380} 417--425 (\textit{Preprint}
  \eprint{hep-ph/9604317})

\bibitem{Radyushkin:1996ru}
Radyushkin A~V 1996 {\em Phys. Lett.\/} {\bf B385} 333--342 (\textit{Preprint}
  \eprint{hep-ph/9605431})

\bibitem{Ji:1996nm}
Ji X~D 1997 {\em Phys. Rev.\/} {\bf D55} 7114--7125 (\textit{Preprint}
  \eprint{hep-ph/9609381})

\bibitem{Collins:1996fb}
Collins J~C, Frankfurt L and Strikman M 1997 {\em Phys. Rev.\/} {\bf D56}
  2982--3006 (\textit{Preprint} \eprint{hep-ph/9611433})

\bibitem{Radyushkin:1997ki}
Radyushkin A~V 1997 {\em Phys. Rev.\/} {\bf D56} 5524--5557 (\textit{Preprint}
  \eprint{hep-ph/9704207})

\bibitem{Ji:1998xh}
Ji X~D and Osborne J 1998 {\em Phys. Rev.\/} {\bf D58} 094018
  (\textit{Preprint} \eprint{hep-ph/9801260})

\bibitem{Collins:1998be}
Collins J~C and Freund A 1999 {\em Phys. Rev.\/} {\bf D59} 074009
  (\textit{Preprint} \eprint{hep-ph/9801262})

\bibitem{Mueller:1998fv}
Mueller D, Robaschik D, Geyer B, Dittes F~M and Horejsi J 1994 {\em Fortschr.
  Phys.\/} {\bf 42} 101 (\textit{Preprint} \eprint{hep-ph/9812448})

\bibitem{Diehl:2003ny}
Diehl M 2003 {\em Phys. Rept.\/} {\bf 388} 41--277 habilitation Thesis
  (\textit{Preprint} \eprint{hep-ph/0307382})

\bibitem{Polyakov:2002yz}
Polyakov M~V 2003 {\em Phys. Lett.\/} {\bf B555} 57--62 (\textit{Preprint}
  \eprint{hep-ph/0210165})

\bibitem{Belitsky:2005qn}
Belitsky A~V and Radyushkin A~V 2005 {\em Phys. Rept.\/} {\bf 418} 1--387
  (\textit{Preprint} \eprint{hep-ph/0504030})

\bibitem{Burkardt:2000za}
Burkardt M 2000 {\em Phys. Rev.\/} {\bf D62} 071503 (\textit{Preprint}
  \eprint{hep-ph/0005108})

\bibitem{Ralston:2001xs}
Ralston J~P and Pire B 2002 {\em Phys. Rev.\/} {\bf D66} 111501
  (\textit{Preprint} \eprint{hep-ph/0110075})

\bibitem{Miller:2007uy}
Miller G~A 2007 {\em Phys. Rev. Lett.\/} {\bf 99} 112001 (\textit{Preprint}
  \eprint{0705.2409})

\bibitem{Burkardt:2002hr}
Burkardt M 2003 {\em Int. J. Mod. Phys.\/} {\bf A18} 173--208
  (\textit{Preprint} \eprint{hep-ph/0207047})

\bibitem{Diehl:2002he}
Diehl M 2002 {\em Eur. Phys. J.\/} {\bf C25} 223--232 (\textit{Preprint}
  \eprint{hep-ph/0205208})

\bibitem{Burkardt:2007sc}
Burkardt M 2007  [hep-ph 0711.1881 ] (\textit{Preprint} \eprint{0711.1881})

\bibitem{Guichon:1998xv}
Guichon P~A~M and Vanderhaeghen M 1998 {\em Prog. Part. Nucl. Phys.\/} {\bf 41}
  125--190 (\textit{Preprint} \eprint{hep-ph/9806305})

\bibitem{Belitsky:2001ns}
Belitsky A~V, Mueller D and Kirchner A 2002 {\em Nucl. Phys.\/} {\bf B629}
  323--392 (\textit{Preprint} \eprint{hep-ph/0112108})

\bibitem{Goeke:2001tz}
Goeke K, Polyakov M~V and Vanderhaeghen M 2001 {\em Prog. Part. Nucl. Phys.\/}
  {\bf 47} 401--515 (\textit{Preprint} \eprint{hep-ph/0106012})

\bibitem{Diehl:1997bu}
Diehl M, Gousset T, Pire B and Ralston J~P 1997 {\em Phys. Lett.\/} {\bf B411}
  193--202 (\textit{Preprint} \eprint{hep-ph/9706344})

\bibitem{Belitsky:2008bz}
Belitsky A~V and Muller D 2009 {\em Phys. Rev.\/} {\bf D79} 014017
  (\textit{Preprint} \eprint{0809.2890})

\bibitem{Guichon:2009}
Guichon P and Vanderhaeghen M 2008 {In preparation}

\bibitem{Radyushkin:1998es}
Radyushkin A~V 1999 {\em Phys. Rev.\/} {\bf D59} 014030 (\textit{Preprint}
  \eprint{hep-ph/9805342})

\bibitem{Vanderhaeghen:1999xj}
Vanderhaeghen M, Guichon P~A~M and Guidal M 1999 {\em Phys. Rev.\/} {\bf D60}
  094017 (\textit{Preprint} \eprint{hep-ph/9905372})

\bibitem{Polyakov:1999gs}
Polyakov M~V and Weiss C 1999 {\em Phys. Rev.\/} {\bf D60} 114017
  (\textit{Preprint} \eprint{hep-ph/9902451})

\bibitem{Petrov:1998kf}
Petrov V~Y {\em et~al.\/} 1998 {\em Phys. Rev.\/} {\bf D57} 4325--4333
  (\textit{Preprint} \eprint{hep-ph/9710270})

\bibitem{Guidal:2004nd}
Guidal M, Polyakov M~V, Radyushkin A~V and Vanderhaeghen M 2005 {\em Phys.
  Rev.\/} {\bf D72} 054013 (\textit{Preprint} \eprint{hep-ph/0410251})

\bibitem{Penttinen:1999th}
Penttinen M, Polyakov M~V and Goeke K 2000 {\em Phys. Rev.\/} {\bf D62} 014024
  (\textit{Preprint} \eprint{hep-ph/9909489})

\bibitem{Diehl:1998kh}
Diehl M, Feldmann T, Jakob R and Kroll P 1999 {\em Eur. Phys. J.\/} {\bf C8}
  409--434 (\textit{Preprint} \eprint{hep-ph/9811253})

\bibitem{Brodsky:2000xy}
Brodsky S~J, Diehl M and Hwang D~S 2001 {\em Nucl. Phys.\/} {\bf B596} 99--124
  (\textit{Preprint} \eprint{hep-ph/0009254})

\bibitem{Boffi:2002yy}
Boffi S, Pasquini B and Traini M 2003 {\em Nucl. Phys.\/} {\bf B649} 243--262
  (\textit{Preprint} \eprint{hep-ph/0207340})

\bibitem{Ji:2006ea}
Ji C~R, Mishchenko Y and Radyushkin A 2006 {\em Phys. Rev.\/} {\bf D73} 114013
  (\textit{Preprint} \eprint{hep-ph/0603198})

\bibitem{Frankfurt:1997ha}
Frankfurt L, Freund A, Guzey V and Strikman M 1998 {\em Phys. Lett.\/} {\bf
  B418} 345--354 (\textit{Preprint} \eprint{hep-ph/9703449})

\bibitem{Freund:2002qf}
Freund A, McDermott M and Strikman M 2003 {\em Phys. Rev.\/} {\bf D67} 036001
  (\textit{Preprint} \eprint{hep-ph/0208160})

\bibitem{Polyakov:2002wz}
Polyakov M~V and Shuvaev A~G 2002  [hep-ph/0207153 ] (\textit{Preprint}
  \eprint{hep-ph/0207153})

\bibitem{Guzey:2005ec}
Guzey V and Polyakov M~V 2006 {\em Eur. Phys. J.\/} {\bf C46} 151--156
  (\textit{Preprint} \eprint{hep-ph/0507183})

\bibitem{Guzey:2008ys}
Guzey V and Teckentrup T 2009 {\em Phys. Rev.\/} {\bf D79} 017501
  (\textit{Preprint} \eprint{0810.3899})

\bibitem{Polyakov:2008aa}
Polyakov M~V and Semenov-Tian-Shansky K~M 2009 {\em Eur. Phys. J.\/} {\bf A40}
  181--198 (\textit{Preprint} \eprint{0811.2901})

\bibitem{Kumericki:2009uq}
Kumericki K and Mueller D 2010 {\em Nucl. Phys.\/} {\bf B841} 1--58
  (\textit{Preprint} \eprint{0904.0458})

\bibitem{Adloff:2001cn}
Adloff C {\em et~al.\/} (H1) 2001 {\em Phys. Lett. B\/} {\bf 517} 47--58
  (\textit{Preprint} \eprint{hep-ex/0107005})

\bibitem{Aktas:2005ty}
Aktas A {\em et~al.\/} (H1) 2005 {\em Eur. Phys. J.\/} {\bf C44} 1--11
  (\textit{Preprint} \eprint{hep-ex/0505061})

\bibitem{Aaron:2007cz}
Aaron F~D {\em et~al.\/} (H1) 2008 {\em Phys. Lett.\/} {\bf B659} 796--806
  (\textit{Preprint} \eprint{0709.4114})

\bibitem{Chekanov:2003ya}
Chekanov S {\em et~al.\/} (ZEUS) 2003 {\em Phys. Lett.\/} {\bf B573} 46--62
  (\textit{Preprint} \eprint{hep-ex/0305028})

\bibitem{Chekanov:2008vy}
Chekanov S {\em et~al.\/} (ZEUS) 2008  (\textit{Preprint} \eprint{0812.2517})

\bibitem{Airapetian:2001yk}
Airapetian A {\em et~al.\/} (HERMES) 2001 {\em Phys. Rev. Lett.\/} {\bf 87}
  182001 (\textit{Preprint} \eprint{hep-ex/0106068})

\bibitem{Airapetian:2006zr}
Airapetian A {\em et~al.\/} (HERMES) 2007 {\em Phys. Rev.\/} {\bf D75} 011103
  (\textit{Preprint} \eprint{hep-ex/0605108})

\bibitem{:2008jga}
Airapetian A {\em et~al.\/} (HERMES) 2008 {\em JHEP\/} {\bf 06} 066
  (\textit{Preprint} \eprint{0802.2499})

\bibitem{:2010mb}
Airapetian A {\em et~al.\/} (HERMES) 2010 {\em JHEP\/} {\bf 06} 019
  (\textit{Preprint} \eprint{1004.0177})

\bibitem{:2009rj}
Airapetian A {\em et~al.\/} (HERMES) 2009 {\em JHEP\/} {\bf 11} 083
  (\textit{Preprint} \eprint{0909.3587})

\bibitem{Seitz:2004kw}
Seitz B (HERMES) 2004 {\em Nucl. Instrum. Meth.\/} {\bf A535} 538--541

\bibitem{Lehmann:1900zz}
 {\em {Hermes Results On Hard-Exclusive Processes And Prospects Using The New
  Recoil Detector}\/} in the Proceedings of 11th International Conference on
  Meson-Nucleon Physics and the Structure of the Nucleon (MENU 2007), Julich,
  Germany, 10-14 Sep 2007

\bibitem{Stepanyan:2001sm}
Stepanyan S {\em et~al.\/} (CLAS) 2001 {\em Phys. Rev. Lett.\/} {\bf 87} 182002
  (\textit{Preprint} \eprint{hep-ex/0107043})

\bibitem{Kivel:2000fg}
Kivel N, Polyakov M~V and Vanderhaeghen M 2001 {\em Phys. Rev.\/} {\bf D63}
  114014 (\textit{Preprint} \eprint{hep-ph/0012136})

\bibitem{Belitsky:2001yp}
Belitsky A~V, Kirchner A, Mueller D and Schafer A 2001 {\em Phys. Lett.\/} {\bf
  B510} 117--124 (\textit{Preprint} \eprint{hep-ph/0103343})

\bibitem{Mecking:2003zu}
Mecking B~A {\em et~al.\/} (CLAS) 2003 {\em Nucl. Instrum. Meth.\/} {\bf A503}
  513--553

\bibitem{Chen:2006na}
Chen S {\em et~al.\/} (CLAS) 2006 {\em Phys. Rev. Lett.\/} {\bf 97} 072002
  (\textit{Preprint} \eprint{hep-ex/0605012})

\bibitem{E00110}
Roblin Y {\em et~al.\/} (the Hall A DVCS Collaboration) 2000 E00-110 jLab
  Experiment E00-110, Deeply Virtual Compton Scattering at 6 GeV
  \urlprefix\url{http://hallaweb.jlab.org/experiment/DVCS/dvcs.pdf}

\bibitem{E03106}
Voutier E {\em et~al.\/} (the Hall A DVCS Collaboration) 2003 E03-106 jLab
  Experiment E03-106, Deeply Virtual Compton Scattering on the Neutron
  \urlprefix\url{http://hallaweb.jlab.org/experiment/DVCS/dvcs.pdf}

\bibitem{Alcorn:2004sb}
Alcorn J {\em et~al.\/} 2004 {\em Nucl. Instrum. Meth.\/} {\bf A522} 294--346

\bibitem{Camsonne:2005th}
Camsonne A 2005 Ph.D. thesis Universit\'e Blaise Pascal, Clermont-Ferrand,
  France

\bibitem{ARS}
{F Feinstein} 2003 {\em Nucl. Instrum. Meth.\/} {\bf A504} 258

\bibitem{MunozCamacho:2006hx}
Munoz~Camacho C {\em et~al.\/} (Jefferson Lab Hall A) 2006 {\em Phys. Rev.
  Lett.\/} {\bf 97} 262002 (\textit{Preprint} \eprint{nucl-ex/0607029})

\bibitem{Mazouz:2007vj}
Mazouz M {\em et~al.\/} (Jefferson Lab Hall A) 2007 {\em Phys. Rev. Lett.\/}
  {\bf 99} 242501 (\textit{Preprint} \eprint{0709.0450})

\bibitem{Ahmad:2006gn}
Ahmad S, Honkanen H, Liuti S and Taneja S~K 2007 {\em Phys. Rev.\/} {\bf D75}
  094003 (\textit{Preprint} \eprint{hep-ph/0611046})

\bibitem{Gockeler:2003jfa}
Gockeler M {\em et~al.\/} (QCDSF) 2004 {\em Phys. Rev. Lett.\/} {\bf 92} 042002
  (\textit{Preprint} \eprint{hep-ph/0304249})

\bibitem{Brommel:2007sb}
Brommel D {\em et~al.\/} (QCDSF-UKQCD) 2007 {\em PoS\/} {\bf LAT2007} 158
  (\textit{Preprint} \eprint{0710.1534})

\bibitem{Hagler:2007xi}
Hagler P {\em et~al.\/} (LHPC) 2008 {\em Phys. Rev.\/} {\bf D77} 094502
  (\textit{Preprint} \eprint{0705.4295})

\bibitem{Thomas:2008ga}
Thomas A~W 2008 {\em Phys. Rev. Lett.\/} {\bf 101} 102003 (\textit{Preprint}
  \eprint{0803.2775})

\bibitem{E07-007}
Camacho C~M {\em et~al.\/} ({the Hall A DVCS Collaboration}) 2007 {JLab
  E07-007} complete Separation of Deeply Virtual Photon and Neutral Pion
  Electroproduction Observables of Unpolarized Protons
  \urlprefix\url{www.jlab.org/exp_prog/proposals/07/E07-007.pdf}

\bibitem{E08-025}
Mazouz M {\em et~al.\/} 2008 {JLab E08-025} {Measurement of the Deeply Virtual
  Compton Scattering cross-section off the neutron}
  \urlprefix\url{www.jlab.org/exp_prog/proposals/08prop.html/PR-08-025.pdf}

\bibitem{Girod:2007jq}
Girod F~X {\em et~al.\/} (CLAS) 2008 {\em Phys. Rev. Lett.\/} {\bf 100} 162002
  (\textit{Preprint} \eprint{0711.4805})

\bibitem{Guidal:2008ie}
Guidal M 2008 {\em Eur. Phys. J.\/} {\bf A37} 319--332 (\textit{Preprint}
  \eprint{0807.2355})

\bibitem{Laget:2007qm}
Laget J~M 2007 {\em Phys. Rev.\/} {\bf C76} 052201 (\textit{Preprint}
  \eprint{arXiv:0708.1250 [hep-ph]})

\bibitem{E06-003}
Burkert V, Elouadrhiri L, Garcon M, Niyazov R, Stepanyan S {\em et~al.\/} ({the
  CLAS Collaboration}) 2006 {JLab E06-003} {Deeply Virtual Compton Scattering
  with CLAS at 6 GeV }
  \urlprefix\url{http://www.jlab.org/exp_prog/proposals/06/PR06-003.pdf}

\bibitem{E05-114}
Biselli A, Elouadrhiri L, Joo K, Niccolai S {\em et~al.\/} ({the CLAS
  Collaboration}) 2005 {JLab E05-114} {Deeply Virtual Compton Scattering at 6
  GeV with polarized target and polarized beam using the CLAS detector}
  \urlprefix\url{http://www.jlab.org/exp_prog/proposals/05/PR05-114.pdf}

\bibitem{Guidal:2010ig}
Guidal M 2010 {\em Phys. Lett.\/} {\bf B689} 156--162 (\textit{Preprint}
  \eprint{1003.0307})

\bibitem{E06-101}
Klein F, Sandorfi A {\em et~al.\/} ({the CLAS Collaboration}) 2006 {JLab
  E06-101} {N-star Resonances in Pseudoscalar-meson photo-production from
  Polarized Neutrons in $\vec{H}\cdot\vec{D}$ and a complete determination of
  the $\gamma n\rightarrow K^0 \Lambda$ amplitude }
  \urlprefix\url{http://www.jlab.org/exp_prog/proposals/06/PR-06-101.pdf}

\bibitem{Berger:2001zb}
Berger E~R, Cano F, Diehl M and Pire B 2001 {\em Phys. Rev. Lett.\/} {\bf 87}
  142302 (\textit{Preprint} \eprint{hep-ph/0106192})

\bibitem{Cano:2003ju}
Cano F and Pire B 2004 {\em Eur. Phys. J.\/} {\bf A19} 423--438
  (\textit{Preprint} \eprint{hep-ph/0307231})

\bibitem{E08-024}
Egiyan H, Girod F~X, Hafidi K, Liuti S, Voutier E {\em et~al.\/} (CLAS) 2008
  {Deeply Virtual Compton Scattering off ${}^4$He} {JLab E08-024}
  \urlprefix\url{{www.jlab.org/exp$\_$prog/proposals/08/PR-08-024.pdf}}

\bibitem{Voutier:2008wu}
Voutier E  {Proceedings of the International Workshop on Nuclear Theory, Rila
  Mountains, Bulgaria, 23-28 Jun 2008} (\textit{Preprint} \eprint{0809.2670})

\bibitem{Fenker:2008zz}
Fenker H~C {\em et~al.\/} 2008 {\em Nucl. Instrum. Meth.\/} {\bf A592} 273--286

\bibitem{DeMasi:2007id}
De~Masi R {\em et~al.\/} (CLAS) 2008 {\em Phys. Rev.\/} {\bf C77} 042201
  (\textit{Preprint} \eprint{0711.4736})

\bibitem{Fuchey:2010}
Fuchey E {\em et~al.\/} (Jefferson Laboratory Hall A)  (\textit{Preprint}
  \eprint{1003.2938})

\bibitem{Gilman:2010}
Gilman R, Holt R and Stoler P 2010

\bibitem{Hadjidakis:2004zm}
Hadjidakis C {\em et~al.\/} (CLAS) 2005 {\em Phys. Lett.\/} {\bf B605} 256--264
  (\textit{Preprint} \eprint{hep-ex/0408005})

\bibitem{Morrow:2008ek}
Morrow S~A {\em et~al.\/} (CLAS) 2009 {\em Eur Phys J\/} {\bf A39} 5--31
  (\textit{Preprint} \eprint{0807.3834})

\bibitem{Goloskokov:2006hr}
Goloskokov S~V and Kroll P 2007 {\em Eur. Phys. J.\/} {\bf C50} 829--842
  (\textit{Preprint} \eprint{hep-ph/0611290})

\bibitem{Goloskokov:2005sd}
Goloskokov S~V and Kroll P 2005 {\em Eur. Phys. J.\/} {\bf C42} 281--301
  (\textit{Preprint} \eprint{hep-ph/0501242})

\bibitem{Guidal:2007cw}
Guidal M and Morrow S 2007 {Exclusive $\rho^0$ electroproduction on the proton
  : GPDs or not GPDs ?} [hep-ph 0711.3743] (\textit{Preprint}
  \eprint{0711.3743})

\bibitem{Laget:2000gj}
Laget J~M 2000 {\em Phys. Lett.\/} {\bf B489} 313--318 (\textit{Preprint}
  \eprint{hep-ph/0003213})

\bibitem{Cassel:1981sx}
Cassel D~G {\em et~al.\/} 1981 {\em Phys. Rev.\/} {\bf D24} 2787

\bibitem{Airapetian:2000ni}
Airapetian A {\em et~al.\/} (HERMES) 2000 {\em Eur. Phys. J.\/} {\bf C17}
  389--398 (\textit{Preprint} \eprint{hep-ex/0004023})

\bibitem{Adams:1997bh}
Adams M~R {\em et~al.\/} (E665) 1997 {\em Z. Phys.\/} {\bf C74} 237--261

\bibitem{Chekanov:2007zr}
Chekanov S {\em et~al.\/} (ZEUS) 2007 {\em PMC Phys.\/} {\bf A1} 6
  (\textit{Preprint} \eprint{0708.1478})

\bibitem{Morand:2005ex}
Morand L {\em et~al.\/} (CLAS) 2005 {\em Eur. Phys. J.\/} {\bf A24} 445--458
  (\textit{Preprint} \eprint{hep-ex/0504057})

\bibitem{Joos:1977tz}
Joos P {\em et~al.\/} 1977 {\em Nucl. Phys.\/} {\bf B122} 365

\bibitem{Santoro:2008ai}
Santoro J~P {\em et~al.\/} (CLAS) 2008 {\em Phys. Rev.\/} {\bf C78} 025210
  (\textit{Preprint} \eprint{0803.3537})

\bibitem{Borissov:2001fq}
Borissov A~B (HERMES) 2001 {\em Nucl. Phys. Proc. Suppl.\/} {\bf 99A} 156--163

\bibitem{Chekanov:2005cqa}
Chekanov S {\em et~al.\/} (ZEUS) 2005 {\em Nucl. Phys.\/} {\bf B718} 3--31
  (\textit{Preprint} \eprint{hep-ex/0504010})

\bibitem{Adloff:2000nx}
Adloff C {\em et~al.\/} (H1) 2000 {\em Phys. Lett.\/} {\bf B483} 360--372
  (\textit{Preprint} \eprint{hep-ex/0005010})

\bibitem{Laget:2004qu}
Laget J~M 2004 {\em Phys. Rev.\/} {\bf D70} 054023 (\textit{Preprint}
  \eprint{hep-ph/0406153})

\bibitem{d'Hose:2004kj}
d'Hose N, Burtin E, Guichon P~A~M and Marroncle J 2004 {\em Eur. Phys. J.\/}
  {\bf A19} Suppl147--53

\bibitem{Moutarde:2009fg}
Moutarde H 2009 {\em Phys. Rev.\/} {\bf D79} 094021 (\textit{Preprint}
  \eprint{0904.1648})

\end{thebibliography}

\end{document}